\pdfoutput=1
\documentclass[prx,eqsecnum,twocolumn,aps,showpacs,amsmath,amssymb,floatfix,longbibliography]{revtex4-1}
\usepackage{graphicx}
\DeclareGraphicsExtensions{.pdf}
\usepackage{amsmath,amssymb,bbold,bm,color}
\usepackage{float}
\usepackage{epstopdf}
\usepackage[dvipsnames]{xcolor}

\newcommand{\bk}{{\bm k}}
\newcommand{\bK}{{\bm K}}
\newcommand{\bq}{{\bm q}}

\newcommand{\bp}{{\bm p}}
\newcommand{\br}{{\bm r}}
\newcommand{\bv}{{\bm v}}
\newcommand{\bu}{{\bm u}}

\newcommand{\bB}{{\bm B}}
\newcommand{\bE}{{\bm E}}
\newcommand{\bb}{{\bm b}}
\newcommand{\be}{{\bm e}}
\newcommand{\bA}{{\bm A}}\newcommand{\ba}{{\bm a}}

\newcommand{\bj}{{\bm j}}
\newcommand{\bsig}{{\bm \sigma}}
\newcommand{\btau}{{\bm \tau}}

\newcommand{\cT}{{\cal T}}

\newcommand{\bee}{\begin{equation}}
\newcommand{\ee}{\end{equation}}

\def\sgn{\mathop{\rm sgn}\nolimits}

\begin{document}

\title{Chiral anomaly from strain-induced gauge fields in Dirac and Weyl semimetals}
%\title{Chiral anomaly without magnetic field from strain engineering in Weyl and Dirac semimetals}

\author{D.I. Pikulin}
\author{Anffany Chen}
\author{M. Franz}
\affiliation{Department of Physics and Astronomy, University of
British Columbia, Vancouver, BC, Canada V6T 1Z1}
\affiliation{Quantum Matter Institute, University of British Columbia, Vancouver BC, Canada V6T 1Z4}

\date{\today}

\begin{abstract} 
Dirac and Weyl semimetals form an ideal platform for testing ideas developed in high energy physics to describe massless relativistic particles. One such  quintessentially field-theoretic idea of the chiral anomaly already resulted in the prediction and subsequent observation of the pronounced negative magnetoresistance in these novel materials for parallel electric and magnetic fields. Here we predict that the chiral anomaly occurs -- and has experimentally observable consequences -- when real electromagnetic fields $\bE$ and $\bB$ are replaced by strain-induced pseudo-electromagnetic fields $\be$ and $\bb$. For example, a uniform pseudomagnetic field $\bb$ is generated when a Weyl semimetal nanowire is put under torsion. In accord with the chiral anomaly equation we predict a negative contribution to the wire resistance proportional to the square of the torsion strength. Remarkably, left and right moving chiral modes are then spatially segregated to the bulk and surface of the wire forming a ``topological coaxial cable''. This produces hydrodynamic flow with potentially very long relaxation time. Another effect we predict is the ultrasonic attenuation and electromagnetic emission due to a time periodic mechanical deformation causing pseudoelectric field $\be$. These novel manifestations of the chiral anomaly are most striking in the semimetals with a single pair of Weyl nodes but also occur in Dirac semimetals such as Cd$_3$As$_2$ and Na$_3$Bi and Weyl semimetals with unbroken time reversal symmetry.
 
\end{abstract}

\date{\today}

\maketitle

\section{Introduction}

Mechanical strain that varies smoothly on the interatomic scale is known to affect the low-energy Dirac fermions  in graphene in a way that is similar to the externally applied magnetic field. More precisely,  strain acts in graphene as a ``chiral''  vector potential that couples to Dirac fermions oppositely in the two valleys $\bK$ and $\bK'$ \cite{guinea2010}. The pseudomagnetic field that arises from this effect in a curved graphene sheet can be larger than 300T,
and has been observed through the spectroscopic measurement of the Landau levels in the seminal experiment on graphene nanobubbles \cite{levy2010}. In terms of their low-energy physics  Weyl and Dirac semimetals \cite{Savrasov2011,burkov2011b,Vafek2014} can be thought of as three dimensional generalization of graphene. The question thus immediately arises whether strain in these materials gives rise to similar effects. Recent theoretical work \cite{cortijo2015} showed that this is indeed the case at least in a simple toy model of a Weyl semimetal with broken time reversal symmetry $\cT$. The authors predicted that the electron-phonon coupling in such a system will  lead to non-zero phonon Hall viscosity, an interesting but notoriously difficult quantity to measure. We consider here the effect of strain in more realistic models relevant to Dirac semimetals Cd$_3$As$_2$ \cite{zhizhun2013,borisenko2014,neupane2014,jeon2014,he2014,liu2014b} and Na$_3$Bi \cite{zhizhun2012,yulin2014,yulin2014b} and the related Weyl semimetals \cite{hasan2015,ding2015,yan2015,chen2015,xu2015}. We describe situations where the strain-induced pseudo electromagnetic fields $\be$ and $\bb$ give rise to new and unusual manifestations of the chiral anomaly  \cite{adler1969,bell1969,nielsen1983} which can be observed by conventional experimental probes such as electrical transport, ultrasonic attenuation  and electromagnetic field emission.

One reason why strain can generate pseudomagnetic fields as large as 300T in graphene \cite{levy2010} lies in its mechanical flexibility: substantial curvature can be achieved without breaking the graphene sheet. This suggests that to probe strain-induced effects in Dirac and Weyl semimetals one should focus on films or wires as these will be much more flexible than bulk crystals. In this work we thus concentrate on these geometries and show that strain leads to phenomena that are both striking and experimentally measurable. We note that high-quality nanowires of Dirac semimetal Cd$_3$As$_2$ have been grown and shown to exhibit giant negative magnetoresistance due to the chiral anomaly \cite{li2015} as well as Aharonov-Bohm oscillations indicative of the protected surface states \cite{wang2016}. These wires bend easily and show mechanical flexibility  that is required to study strain related phenomena. We also discuss consequences of lattice distortions caused by sound waves (phonons). These can be used to study the above phenomena in crystalline flakes and films which are readily available for nearly all known Dirac and Weyl materials.

\begin{figure*}[t]
\includegraphics[width = 15.5cm]{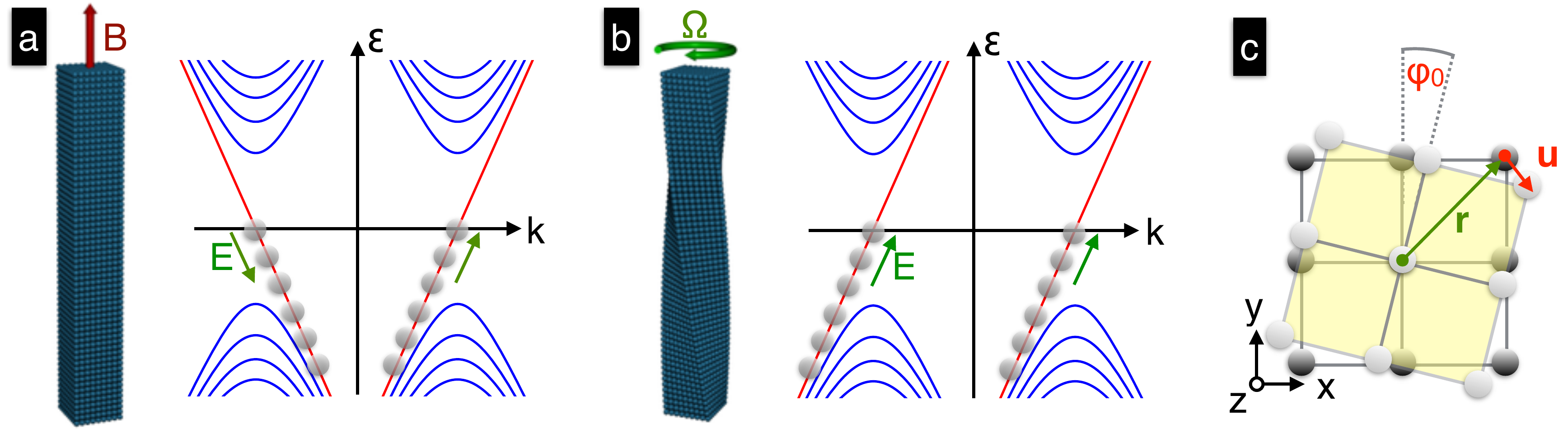}
\caption{Electron excitation spectra in a Weyl semimetal in the presence of a) magnetic field $\bB$ and b) pseudomagnetic field $\bb$ generated  by a torsional deformation. Parallel electric field $\bE$ produces a charge density imbalance in case( a) while it appears to produce excess total charge density in case (b). Panel (c) illustrates the displacement field $\bu$ in the presence of torsion. Consecutive layers of the crystal are rotated by relative angle $\varphi_0=\Omega(L/a)$.
}\label{fig1}
\end{figure*}
Our results can be most easily understood by thinking about the simplest Weyl semimetal with a single pair of Weyl points \cite{burkov2011} although many aspects translate to more complicated Weyl and Dirac semimetals. The low-energy effective theory is then defined by the Hamiltonian $H=\int d^3r\Psi^\dagger_\br h(\br)\Psi_\br$ where $\Psi^\dagger_\br=(\psi^\dagger_{\br,R},\psi^\dagger_{\br,L})$ and
\bee\label{h1}
h=v\chi^z\bsig\cdot(\bp-e\bA-\chi^z e\ba)-\mu.
\ee
Here $\psi^\dagger_{\br,R/L}$ represent  two-component right and left handed Weyl fermion creation operators, $\chi^z=\pm 1$ labels the chirality of the two Weyl nodes, $\bsig$ is a vector of Pauli matrices in the pseudospin space and $\bp=-i\hbar\nabla$.  $\bA$ and $\ba$ denote gauge potentials of the ordinary EM and the chiral field, respectively. We explain below the origin of the chiral fields by considering the effect of the elastic strain in the full lattice model that underlies the low-energy theory (\ref{h1}). Specifically, using the method developed in Refs.\ \cite{cortijo2015,shapourian2015}, we find that (i) a uniform pseudomagnetic field $\bb=\nabla\times \ba$ directed along the axis of the wire $\hat{z}$  is generated by applying static torsion as indicated in Fig. \ref{fig1}b while (ii) pseudoelectric field $\be=-{1\over c}\partial_t\ba$, also along $\hat{z}$, is obtained by dynamically stretching and compressing the sample. 

Consequences of the strain-induced gauge fields can be most easily deduced from the chiral anomaly equations \cite{adler1969,bell1969,nielsen1983} which take the following form  when both ordinary and chiral EM fields are present \cite{qi2013}:
\begin{eqnarray}
\partial_t\rho_5+\nabla\cdot\bj_5&=&{e^2\over 2\pi^2\hbar^2 c}(\bE\cdot\bB+\be\cdot\bb), \label{an1} \\
\partial_t\rho+\nabla\cdot\bj&=&{e^2\over 2\pi^2\hbar^2 c}(\bE\cdot\bb+\be\cdot\bB).
\label{an2}
\end{eqnarray}
Here $\rho$ and $\rho_5$ are the total electron and chiral density, respectively, $\bj$ and $\bj_5$ are the corresponding current densities. Chiral density $\rho_5=\rho_R-\rho_L$ refers to the difference between the charge densities associated with the right- and left-handed Weyl points.

The first equation (\ref{an1}) is most commonly associated with the chiral anomaly, and expresses non-conservation of the chiral charge in the presence of aligned EM or pseudo-EM fields. Physically, this can be understood  as pumping of charge from one Weyl point to the other -- the chiral magnetic effect \cite{nielsen1983}. It is this phenomenon that  underlies the anomalous negative magnetoresistance \cite{fukushima2008,son2013,burkov2015} that has been recently observed in a variety of Weyl and Dirac semimetals \cite{kim2013,huang2015,ong2015,valla2016,jia2016}. 

The second anomaly equation (\ref{an2}) only occurs when both ordinary and pseudo-EM fields are present. It expresses an apparent charge density non-conservation, which is the focus of the present work. In a real solid charge density is of course strictly conserved and Eq.\ (\ref{an2}) therefore must be interpreted with caution. 
%We will see that it nevertheless correctly describes what happens at low energies in the bulk of the system.  The total charge is conserved when all quantum states and boundaries of the system are included, as it must be.
 We will show that Eq.\ (\ref{an2}) can be understood  as pumping of charge between the bulk and the boundary of the system. Such pumping only occurs when either $\bb$ or $\be$ fields are present and furnishes a novel manifestation of the chiral anomaly in a strained crystal.

To develop some intuition for the chiral anomaly let us consider the Hamiltonian (\ref{h1}) in the presence of a static uniform (pseudo)magnetic field. We  begin with the ordinary magnetic field $\bB=B\hat{z}$. The solution of the corresponding Schr\"odinger equation $h\Phi=\epsilon\Phi$ is well known and consists of the set of Dirac Landau levels with energies \cite{nielsen1983}
\bee \label{lan1}
\epsilon_{n}(k)=\pm \hbar v\sqrt{k^2+2n{e|B|\over \hbar c}}, \ \ \ n=1,2,\dots,
\ee
for each Weyl fermion. There is also one chiral $n=0$ level per valley with $\epsilon_{0}(k)=\chi^z \sgn(B) \hbar v k $. If a parallel electric field $\bE =E\hat{z}$ is now applied to the system then the electron momenta begin to evolve according to the semiclassical equation of motion  $k(t)=k(0)-eEt/\hbar$. Because of the existence of the two chiral branches in the spectrum this leads to charge pumping between the two  Weyl points, as illustrated in Fig.\ \ref{fig1}a, at a rate consistent with the chiral anomaly equation (\ref{an1}). The key point here is that in a real solid where the  Hamiltonian is defined on the lattice the two chiral branches are connected away from the Weyl points and the chiral anomaly equation simply describes the semiclassical evolution of the electron states through the Brillouin zone \cite{nielsen1983}. In the presence of relaxation processes a steady state non-equilibrium distribution of electrons with nonzero chiral density $\rho_5$ is obtained which is responsible for the anomalous $\sim B^2$ contribution to the magnetoresistance.

Now consider the effect of the chiral magnetic field $\bb=b\hat{z}$.  The solution consists of the same Dirac Landau levels Eq.\ (\ref{lan1}) but the $n=0$ levels now disperse in the same direction for the two Weyl points, $\epsilon_{0}(k)=\sgn(b) \hbar v k $, as illustrated In Fig.\ \ref{fig1}b. Now if a parallel electric field $\bE =E\hat{z}$ is applied to the system we see that the charge density seemingly begins to change. Since the total charge is conserved this extra charge density must come from somewhere.  We will demonstrate below that it comes from the edge of the system. Indeed this is plausible if we note that the energy spectrum sketched in  Fig.\ \ref{fig1}b does not represent a legitimate dispersion of a lattice system which, due to the periodicity of the energy bands in the momentum space, must exhibit the same number of left and right moving modes. Since the Landau levels are the correct eigenstates in the bulk we conclude that the missing left moving modes must exist at the boundary. Our numerical simulations of a lattice model below indeed confirm this conclusion. Thus, in the presence of $\bb$ and $\bE$ the chiral anomaly can be understood as pumping of charge between the bulk and the edge of the system. The effects of nonzero $\be\cdot\bB$ {and $\be \cdot \bb$} terms are more subtle, as they involve relaxational dynamics, but can be understood from similar arguments. {Indeed, the difference between the effects lies in the directions of magnetic and electric fields as applied to the two Weyl cones.} These effects and their experimental consequences constitute the main result of the paper. 

Several interesting observation follow from the above discussion. First, we conclude that electric transport in a twisted Weyl semimetal wire will be highly unusual because the right-moving modes occur in the bulk whereas the left-moving modes are localized near the boundary. (More precisely we may say that there is a net imbalance between the number of left and right moving modes in the bulk and at the boundary.) Since the left and right moving modes are spatially segregated one expects backscattering to be suppressed in such wires giving rise to anomalously long mean free paths. In addition, transport will sensitively depend on the applied torsion, giving rise to the new chiral torsional effect (CTE) that we describe in detail below. Second, we will see that charge transfer between the bulk and the boundary leads to interesting effects when time-dependent $\be$ field is generated e.g.\ by driving a longitudinal sound wave through the crystal when $\bB$ field is also present. Such a sound wave will experience an anomalous attenuation that can be attributed to the chiral anomaly.  It will also produce charge density oscillations in the crystal that can be observed through electric field measurement outside the sample. Third, the chiral anomaly can be observed even in the complete absence of real EM fields when the crystal is put simultaneously under torsion and time-periodic uniaxial strain. Then nonzero $\be\cdot\bb$ term is generated and according to Eq.\ (\ref{an1}) the chiral charge fails to be conserved. We argue that this has observable consequences for sound attenuation in the crystal.   

Finally, we note the similarity of the second chiral anomaly equation  \eqref{an2} to the equation of parity anomaly in rotating liquid He \cite{volovik2003}. Though the anomaly equations are similar in the two systems (missing the $\be\cdot \bB$ term in the Helium case), the suggested experimental systems and manifestations are very different -- we propose torsion not rotation, and transport not force measurement.

%%%%%%%%%%%%%
\section{Gauge fields from strain in the lattice model of ${\rm Cd}_3{\rm As}_2$ and ${\rm Na}_3{\rm Bi}$ }

We now proceed to justify the above claims by detailed model calculations. For simplicity and concreteness we adopt a specific model describing the low-energy degrees of freedom in the Dirac semimetal Cd$_3$As$_2$. The model captures the band inversion of the atomic Cd-$5s$ and As-$4p$ levels near the $\Gamma$ point. In the basis of the relevant spin-orbit coupled states $|P_{3\over 2},{3\over 2}\rangle$, $|S_{1\over 2},{1\over 2}\rangle$,   $|S_{1\over 2},-{1\over 2}\rangle$ and  $|P_{3\over 2},-{3\over 2}\rangle$ it is defined by a $4\times 4$ matrix Hamiltonian \cite{zhizhun2013}
\bee\label{h2}
H(\bk)=\epsilon_0(\bk)+
\begin{pmatrix}
M(\bk) & Ak_- & 0 & 0\\
Ak_+& -M(\bk) & 0 & 0 \\
0 & 0 & -M(\bk) &-Ak_- \\
0 & 0 & -Ak_+ & M(\bk)
\end{pmatrix}.
\ee
Here $\epsilon_0(\bk)=C_0+C_1k_z^2+C_2(k_x^2+k_y^2)$, $k_\pm=k_x\pm ik_y$, and 
 $M(\bk)=M_0+M_1k_z^2+M_2(k_x^2+k_y^2)$. Parameters $C_j$, $A$ and $M_j$ follow from the $\bk\cdot\bp$ expansion of the first principles calculation \cite{zhizhun2013} and are summarized in Appendix A. We note that $H(\bk)$ (with different parameters) also describes Dirac semimetal Na$_3$Bi \cite{zhizhun2012}.  

The low-energy spectrum of the model (\ref{h2}) consists of a pair of Dirac points located at 
\bee\label{Q}
\bK_\eta=(0,0,\eta Q), \ \ \ Q=\sqrt{-M_0/M_1},
\ee
where $\eta=\pm$ is the valley index.
 The model respects time reversal symmetry $\cT=i\sigma^y\tau^xK$, where $K$ denotes complex conjugation and $\bsig$, $\btau$ are Pauli matrices in spin and orbital space, respectively. 
{  $\cT$ maps the upper diagonal (spin up) block $h(\bk)$ of $H(\bk)$ onto the lower diagonal (spin down)  block $-h(\bk)$ and vice versa. 

Since spin up and spin down blocks are effectively decoupled in the model Hamiltonian (\ref{h2})  we can analyze them separately. It is easy to see that 
each diagonal block taken in isolation can be regarded as describing a minimal $\cT$-breaking Weyl semimetal with one pair of Weyl nodes located at $\bK_\pm$. In the following we will often focus our discussion on the spin up block of Hamiltonian (\ref{h2}) and  refer to it as ``${1\over 2}$-Cd$_3$As$_2$'' model. Once we have understood the physics of this ${1\over 2}$-Cd$_3$As$_2$ model  it will be straightforward  to deduce the behavior of the actual Cd$_3$As$_2$ by simply adding a time-reversal conjugate set of states to the results obtained for ${1\over 2}$-Cd$_3$As$_2$.}
We emphasize that although ${1\over 2}$-Cd$_3$As$_2$ model taken on its own does not describe any specific real material the results we report for this model are relevant to a broad class of Weyl semimetals with broken $\cT$ such as the Burkov-Balents layered heterostructure \cite{burkov2011} and more recently proposed magnetic Weyl materials \cite{liu2016,tang2016}. We will explain in detail how these results apply to $\cT$-preserving Weyl and  Dirac semimetals. 

For many considerations and for numerical calculations it will be useful to regularize the model defined by Eq.\ (\ref{h2}) on a lattice. 
{  Although real Cd$_3$As$_2$ crystal has a complex structure with 40 atoms per unit cell, Ref.\  \cite{zhizhun2013} showed that its low-energy physics can be well described by an effective tight binding model with $s$ and $p$ orbitals on vertices of the tetragonal lattice and lattice constants $a_x,a_y=3.0$\AA \ and $a_z=5.0$\AA. Here we simplify the model one step further and assume a simple cubic lattice with a lattice constant $a$. We checked that this leads to only minor deviations from the tetragonal model of Ref.\ \cite{zhizhun2013}.  We construct the tight-binding model for Cd$_3$As$_2$, as further explained in Appendix A, such that in the vicinity of the $\Gamma$ point it matches the $\bk\cdot\bp$ Hamiltonian  (\ref{h2}) to the leading order in the expansion in small $ak$. For quantitative estimates we use $a=4$\AA \  while in the numerics we use larger values of $a$ as this will allow us to simulate systems of sufficient size with the available computational resources. This does not affect the qualitative features of the physics we wish to describe.} The Cd$_3$As$_2$ Hamiltonian regularized on the lattice thus becomes
\bee \label{h03}
 H^{\rm latt} =\epsilon_\bk+
\begin{pmatrix}
h^{\rm latt} & 0 \\
0 & -h^{\rm latt}
\end{pmatrix},
\ee
where $\epsilon_\bk$ is the lattice version of $\epsilon_0(\bk)$  given in Appendix A while 
\bee \label{h3}
h^{\rm latt}(\bk)=m_\bk\tau^z+\Lambda(\tau^x\sin{ak_x}+\tau^y\sin{ak_y}).
\ee
Here $m_\bk=t_0+t_1\cos{ak_z} +t_2(\cos{ak_x}+\cos{ak_y})$ and $t_0=M_0+2(M_1+2M_2)/a^2$, $t_{1/2}=-2M_{1/2}/a^2$, $\Lambda=A/a$. 

The Hamiltonian (\ref{h3}) exhibits a single pair of Weyl nodes at $\bK_\eta=(0,0,\eta Q)$ and $Q$ given by $\cos(aQ)=-(t_0+2t_2)/t_1$ which coincides with Eq.\ (\ref{Q})  in the limit $aQ\ll 1$. In the vicinity of the nodes we can expand  $h^{\rm latt}(\bK_\pm+\bq)$ in $\bq$ to obtain the Weyl Hamiltonian
\bee \label{h3c}
h_{\eta}(\bq)=\hbar v_{\eta}^j\tau^j q_j,
\ee
with the velocity vector 
\bee \label{h3cc}
\bv_{\eta}=\hbar^{-1}a(\Lambda, \Lambda, -\eta t_1 \sin{aQ}).
\ee
 For Cd$_2$As$_3$ parameters and a physical lattice constant $a=4$\AA \ this gives $\hbar\bv_{\eta}=(0.89,0.89,-1.24\eta)$eV\AA.
 From Eq.\ (\ref{h3cc}) we can read off the chiral charge of the Weyl node located at valley $\eta$
\bee \label{h3d}
\chi_{\eta}=\sgn(v^x_{\eta}v^y_{\eta}v^z_{\eta})=-\eta.
\ee

\begin{figure}[t]
\includegraphics[width = 7.5cm]{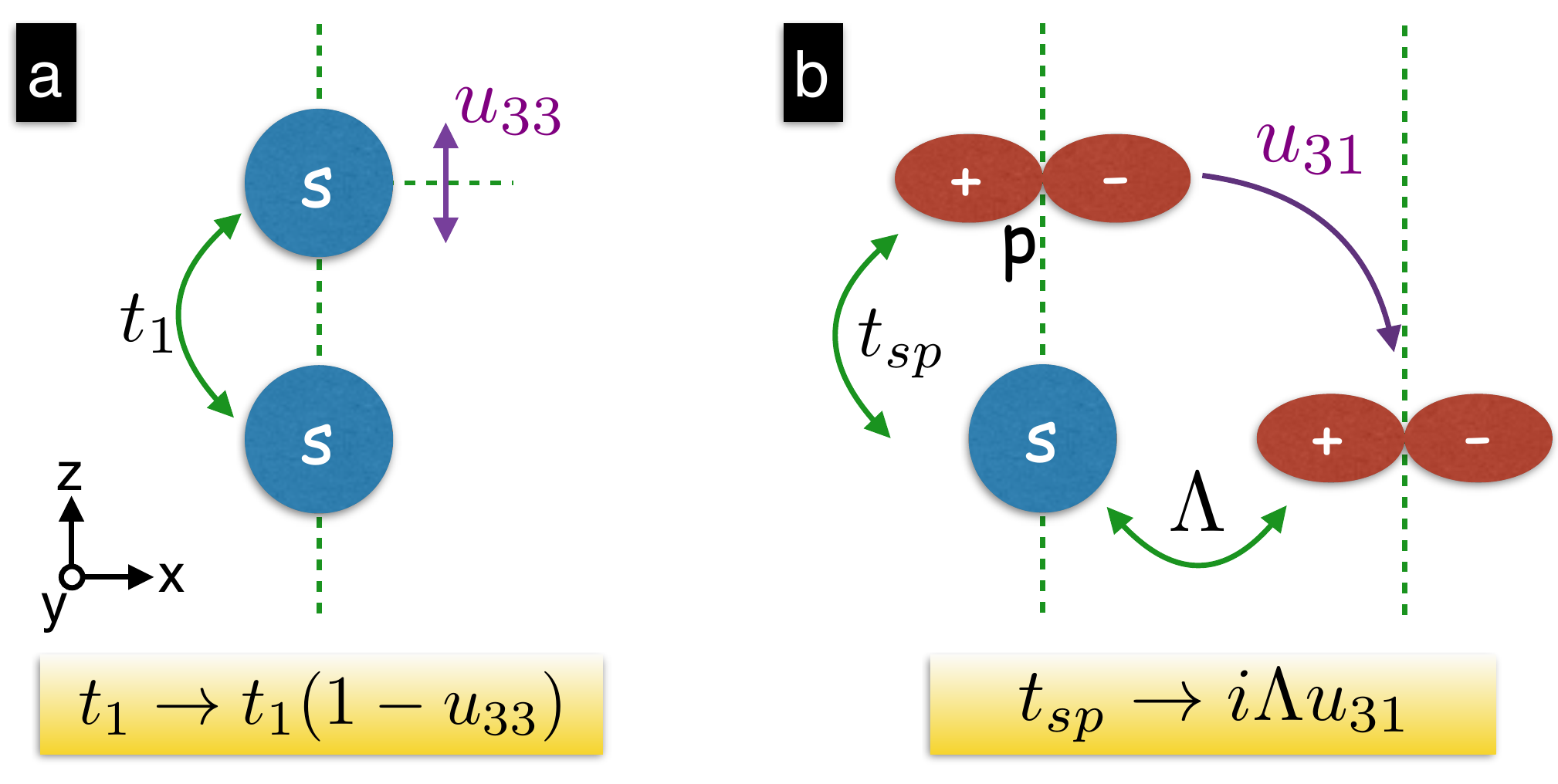}
\caption{  The effect of strain on the hopping amplitudes in the tight binding model. a) Unidirectional strain along the $z$ axis simply changes the distance between the  neighboring orbitals leading to the modification of the hopping amplitude $t_1$ that is linear in $u_{33}$ to leading order in small displacement. b) Torsional strain changes the relative orientation of the orbitals and brings about hopping amplitudes that are disallowed by symmetry in the unstrained crystal, such as $t_{sp}$. The corresponding mathematical expression encodes the expectation that $t_{sp}$ would become equal to $\Lambda$ if the $p$ orbital were displaced all the way to the horizontal position.  In the real material one of course expects Eq.\ (\ref{h4}) to be valid only for displacements small compared to the lattice parameter $a$.
}\label{fig77}
\end{figure}

The effect of strain on the lattice Hamiltonian (\ref{h3}) is implemented using the method developed in Refs.\ \cite{cortijo2015,shapourian2015}.  The key observation is that certain tunneling amplitudes that are prohibited by symmetry in the unstrained crystal become allowed when the strain is applied because of the displacement and rotation of the relevant orbitals in the neighboring atoms. For our purposes the most important modification of the Hamiltonian (\ref{h3}) comes from the replacement of the hopping amplitude along the $\hat{z}$-direction
\cite{cortijo2015,shapourian2015}
\bee \label{h4}   
t_1\tau^z\to t_1(1-u_{33})\tau^z+i\Lambda\sum_{j\neq 3}u_{3j}\tau^j,
\ee
where $u_{ij}={1\over 2}(\partial_iu_j+\partial_ju_i)$ is the symmetrized strain tensor and $\bu=(u_1,u_2,u_3)$ represents the displacement vector. 
{  The physics of  Eq.\ (\ref{h4}) has been discussed at length in Ref.\ \cite{shapourian2015} and is easy to understand intuitively by inspecting the two examples of strain configurations given in Fig.\ \ref{fig77}. The first term in Eq.\ (\ref{h4}) reflects the change in  the hopping amplitude $t_1$ between two like orbitals
(Fig.\ \ref{fig77}a)  when the distance $d$ between the neighboring atoms changes due to strain. The amplitude depends exponentially on $d$ but for small strain it can be expanded to leading order in the atomic displacements which leads to a correction proportional to $u_{33}$. The second term describes generation of hopping processes along the  $\hat{z}$-direction between different orbitals (Fig.\ \ref{fig77}b)   which are prohibited in the unstrained crystal due to their $s$- and $p$- symmetry. The underlying mechanism is outlined  in the caption of Fig.\ \ref{fig77}.
}

As a simple example consider stretching the crystal along the  $\hat{z}$-direction. This is represented by a displacement field $\bu=(0,0,\alpha z)$ where $\alpha=\Delta L/L$ measures the elongation of the crystal. The only nonzero component of the strain tensor is $u_{33}=\alpha$ and Eq.\  (\ref{h4}) thus gives $t_1\to t_1(1-\alpha)$.  It is easy to deduce that for small $\alpha$ this changes the value of $Q\to Q-\alpha Q/(aQ)^2$ thus moving the Weyl nodes closer together or farther apart depending on the sign of $\alpha$ . We see that stretching the crystal has the same effect on the Weyl fermions as the $z$-component of the chiral gauge field $\ba$.

More generally elastic distortion expressed through Eq.\ (\ref{h4}) generates additional terms in the lattice Hamiltonian (\ref{h3}) of the form  
\begin{equation}\label{h5}
\delta h^{\rm latt}(\bk)=-t_1u_{33}\tau^z\cos{ak_z}
+\Lambda(u_{13}\tau^x-u_{23}\tau^y)\sin{ak_z}.
\end{equation}
Expanding again in the vicinity of $\bK_\pm$ we obtain the linearized Hamiltonian of the distorted crystal
\bee \label{h6}
h_{\eta}(\bq)= v_{\eta} ^j\tau^j\left(\hbar q_j-\eta{e\over c}a_j\right),
\ee
where the gauge potential is given by
\bee \label{h7}
\ba=-{\hbar c\over ea}\bigl(u_{13}\sin{aQ},u_{23}\sin{aQ},u_{33}\cot{aQ}\bigr).
\ee
For $aQ\ll 1$ we may approximate $\sin{aQ}\simeq aQ\simeq a\sqrt{-M_0/M_1}$ and $\cot{aQ}\simeq 1/aQ$.

We thus conclude that in a Weyl semimetal with nodes located on the $k_z$ axis components $u_{j3}$ of the strain field act  on the low-energy fermions as a gauge potential. $\ba$ represents a chiral gauge field because it couples with the opposite sign to the Weyl fermions with different chirality $\chi$. 

We saw above that $a_3\sim u_{33}$ can be generated by stretching or compressing the crystal along its $\hat{z}$ axis. Time-dependent distortion of this type will thus produce a pseudoelectric field $\be=-{1\over c}\partial_t\ba$ directed along  $\hat{z}$. In combination with an applied magnetic field $\bB\parallel \hat{z}$ this will generate nonzero $\be\cdot\bB$ term and, as we discuss below,  allow to test the second chiral anomaly equation (\ref{an2}). It is also possible to generate the pseudomagnetic field by applying torsion to the crystal prepared in a wire geometry. To see this consider the displacement field $\bu$ that results from twisting a wire-shaped crystal of length $L$ by angle $\Omega$. As illustrated in Fig.\ \ref{fig1}c we have 
\bee \label{h8}
\bu=\Omega{z\over L}(\br\times \hat{z}),
\ee
where $\br$ denotes the position relative to the origin located on the axis of the wire. Nonzero components of the strain field are $u_{13}=(\Omega/2L)y$ and  $u_{23}=-(\Omega/2L)x$. Via Eq.\ (\ref{h7}) we then get  the pseudomagnetic field 
\bee \label{h9}
\bb=\nabla\times\ba=b_0\hat{z}, \ \ \ b_0=\Omega {\hbar c \over  2Lae}\sin{aQ}.
\ee

To close this Section we estimate the magnitude of the strain-induced field $\bb$ that can be achieved in a typical Cd$_3$As$_2$ nanowire described in Ref.\ \cite{li2015}. We consider a cylindrical wire with a diameter $d=100$nm, length $L=1\mu$m and lattice parameter $a=4$\AA. Eq.\ (\ref{Q}) gives $Q=0.033$\AA$^{-1}$ so the the condition $aQ\ll 1$ is satisfied and we may expand the sine in Eq.\ (\ref{h9}). Recalling further that $\Phi_0=hc/e\simeq 4.12\times 10^5$T\AA$^2$ we find $b_0\approx 1.8\times 10^{-3}$T per angular degree of twist. The maximum attainable field strength in a given wire will depend on how much torsion can the wire sustain before breaking. While we were unable to find any data on the mechanical properties of Cd$_3$As$_2$ we note that Ref.\ \cite{li2015} characterized the nanowires as ``greatly flexible''. We take this to imply that they can withstand substantial torsion. Based on this, a twist angle $\Omega\simeq 180^{\rm o}$ would appear sustainable and will produce $b_0\approx 0.3$T.  For the wire under consideration such a twist translates to a maximum displacement at the outer radius of the wire of about 0.3\AA  \ between the neighboring atoms, or about 8\% of the unit cell. Because the maximum twist angle is limited by the maximum distortion higher effective fields can be achieved in thinner wires.    

%%%%%
\section{Lattice model results}
\begin{figure*}[t]
\includegraphics[width = 17.5cm]{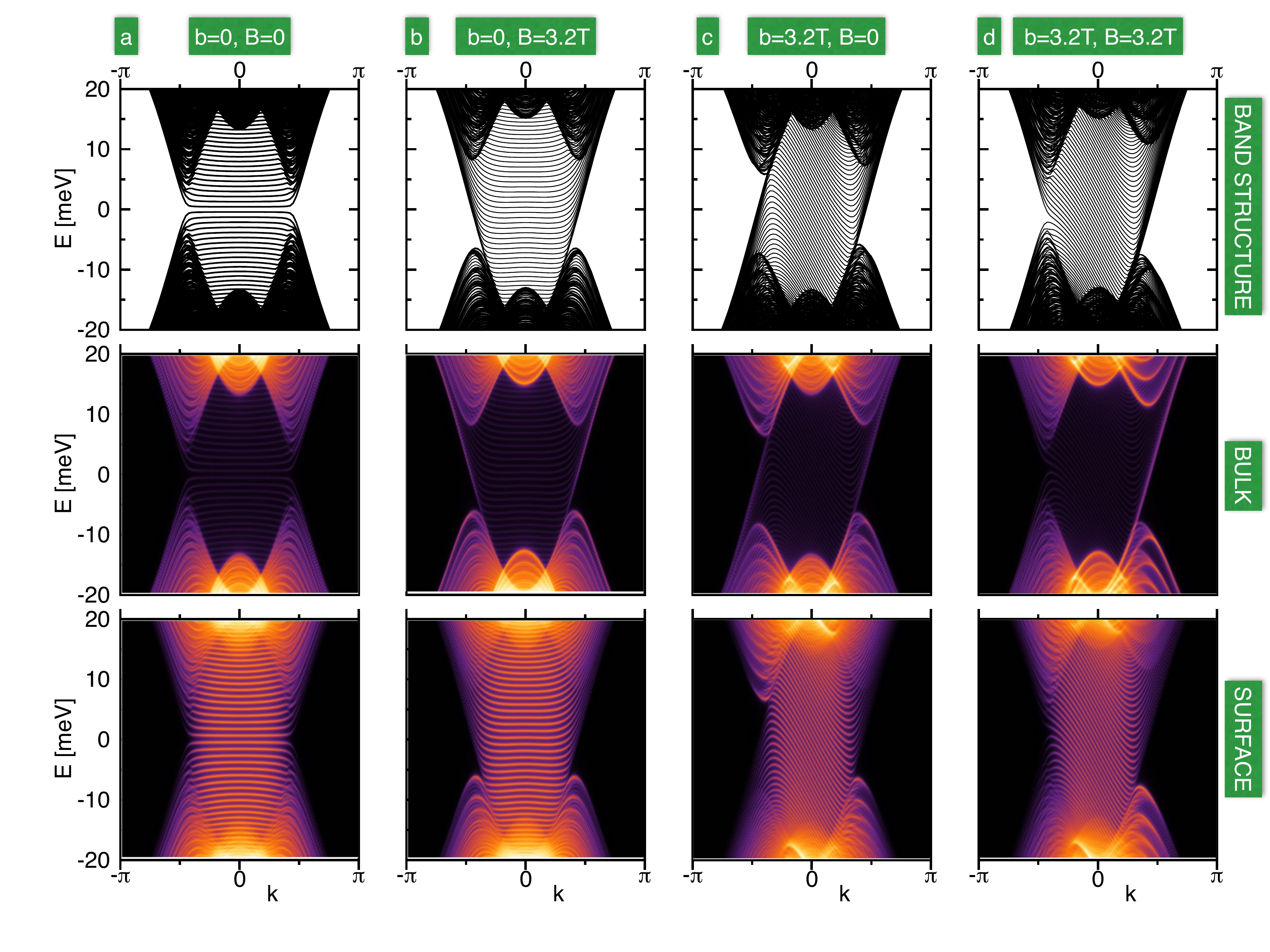}
\caption{Tight-binding model simulations of a Weyl semimetal wire under torsional strain and applied magnetic field $\bB=\hat{z}B$. Top row of figures shows the band structure of the lattice Hamiltonian defined by Eqs.\ (\ref{h3}) and (\ref{h5}) {computed for ${1\over 2}$-Cd$_3$As$_2$ model parameters}, for a wire with a rectangular cross section of $30\times 30$ sites and a lattice constant $a=40$\AA. (We use larger lattice constant here and in subsequent simulations than in real Cd$_3$As$_2$ in order to be able to model nanowires and films of realistic cross sections with available computational resources. Note that this does not affect the physics at low energies because the lattice Hamiltonian is designed to reproduce the relevant $\bk\cdot\bp$ theory independent of $a$.)   
 Open boundary conditions are imposed along $x$ and $y$, periodic along $z$.  Parameters appropriate for Cd$_3$As$_2$ are used. Middle and bottom rows show spectral functions $A^{\rm bulk}(\bk,\omega)$ and  $A^{\rm surf}(\bk,\omega)$. The former is obtained by averaging the full spectral function  $A_{j}(\bk,\omega)$  over sites $j$ in the central $10\times 10$ portion of the wire while the latter averages over  the sites located at the perimeter of the wire. The torsion applied in columns c and d corresponds to the maximum displacement at the perimeter of $0.5 a$, or $\varphi_0\simeq 2^{\rm o}$ between consecutive layers. 
}\label{fig2}
\end{figure*}
To further confirm the validity of the analytical results presented in the previous Sections we carried out extensive numerical simulations of the lattice Hamiltonian (\ref{h3}) in the presence of magnetic field $\bB$ as well as torsional and unidirectional strain implemented via Eq.\ (\ref{h5}). Magnetic field was implemented through the usual Peierls substitution. Our results below indeed validate the general concepts discussed above and illustrate them in a concrete setting of a lattice model relevant to Cd$_3$As$_2$ and Na$_3$Bi.

\subsection{Pseudomagnetic field $\bb$ from torsion}

We start by studying a wire grown along the crystallographic $z$ axis in the presence of magnetic field $\bB=\hat{z}B$ and torsion. 
Representative results are displayed in Fig.\ \ref{fig2}. For simplicity and ease of interpretation we used here parameters appropriate for Cd$_3$As$_2$ (summarized in Appendix A), neglecting terms in $\epsilon_\bk$. We have verified that substantially similar results are obtained when $\epsilon_\bk$ is retained as well as for parameters appropriate for Na$_3$Bi. These results are given in Appendix A.

Column (a) in Fig.\ \ref{fig2} shows the spectrum of an unstrained wire in zero field. Gapless Weyl points are apparent at $k=\pm Q$ and are connected by surface states that originate from the Fermi arcs, expected to occur in the surface of a Weyl semimetal. Spectral functions computed in the bulk,  $A^{\rm bulk}(\bk,\omega)$, and at the surface, 
$A^{\rm surf}(\bk,\omega)$, confirm this identification of bulk and surface electron states. Column (b) exhibits our results for an unstrained wire in magnetic field $B=3.2$T along the axis of the wire. As expected on the basis of  arguments that led to Fig.\ \ref{fig1}a, we observe at low energies a pair of left and right moving chiral modes. These originate from the $n=0$ Landau level and occur in the bulk of the sample. We also observe that the surface states remain largely unaffected by the field.

Our main finding is illustrated in column (c). Torsional strain applied to the wire produces two {\em right moving} chiral modes that are localized in the bulk of the sample as evidenced by  $A^{\rm bulk}(\bk,\omega)$. The bulk spectrum has the structure depicted in Fig.\ \ref{fig1}b expected to occur in the presence of the {\em chiral} magnetic field $\bb$. We are thus led to identify the torsional strain with the chiral vector potential $\ba$. Surface states discernible in the corresponding $A^{\rm surf}(\bk,\omega)$ are seen to compensate for the bulk band structure by providing the required left moving chiral modes.

Column (d)  shows the spectrum for the case when the strength of $\bB$ is chosen to exactly equal $\bb$. As a result, vector potentials $\bA$ and $\ba$ add in one Weyl point but cancel in the other. The resulting  spectrum exhibits a set of right moving bulk chiral modes present in only one of the two Weyl points.  This establishes the complete equivalence of the real magnetic field $\bB$ and the strain-induced pseudomagnetic field $\bb$ insofar as their action on the low-energy Weyl fermions is concerned. 

We note that pseudomagnetic field $b\simeq 3.2$T indicated in Fig.\ \ref{fig1} is larger than the maximum achievable field in the realistic Cd$_3$As$_2$ wire estimated in the previous Section. This is because for clarity we employed here larger torsion (resulting in the maximum displacement of about half the lattice spacing) than can likely be sustained in a real wire. For weaker torsion strengths the effect remains qualitatively unchanged  but becomes less clearly visible in the numerical data for system sizes that are accessible to our simulations.

Results presented in Fig.\ \ref{fig2} pertain to a Weyl semimetal described by Hamiltonian  (\ref{h3}) but are easily extended to Cd$_3$As$_2$ as long as we continue neglecting the particle-hole symmetry breaking term $\epsilon_\bk$. In this limit spectra for Cd$_3$As$_2$ are obtained by simply superimposing bands $E_\bk$ and $-E_\bk$ shown in Fig.\ \ref{fig2}  or by forming spectral functions $A(\bk,\omega)+A(\bk,-\omega)$. Full spectra, including the p-h breaking terms are more complicated but show the same qualitative features. Some relevant examples are given in Appendix A.

\subsection{Pseudoelectric field $\be$ from unidirectional strain}

\begin{figure*}[t]
\includegraphics[width = 17.5cm]{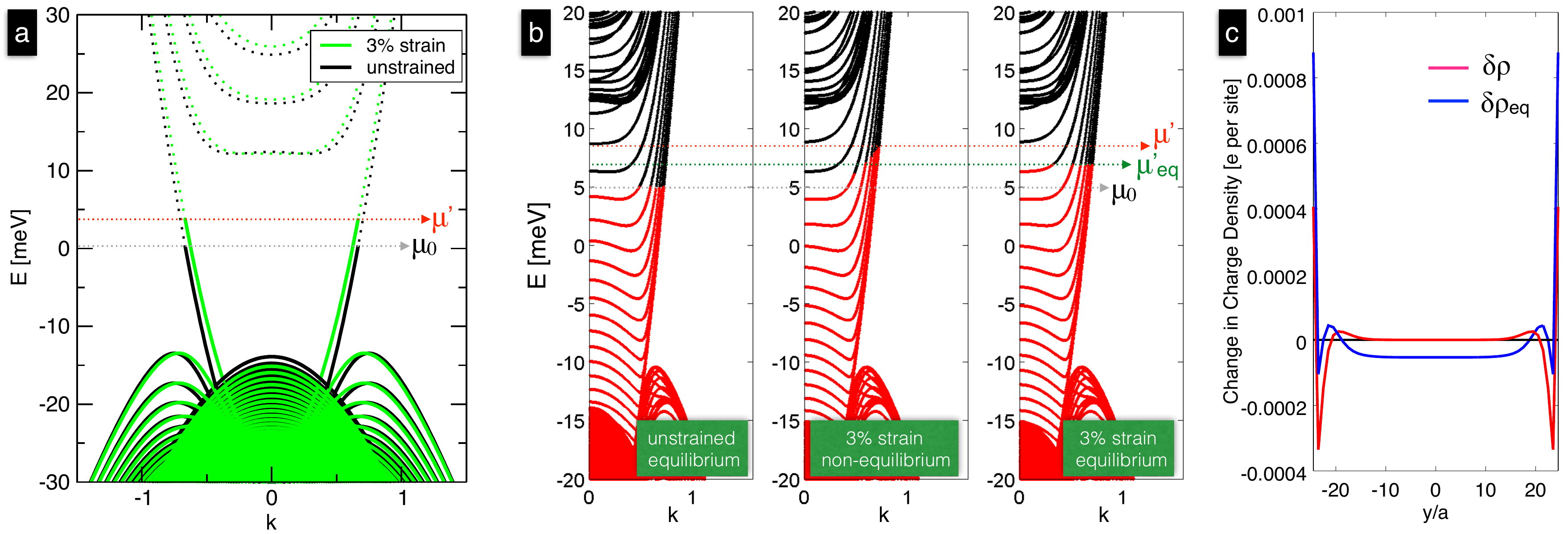}
\caption{Tight-binding model simulations of a Weyl semimetal under applied magnetic field $\bB=\hat{z}B$ and unidirectional strain. Parameters for Cd$_3$As$_2$ listed in Appendix A are used in all panels. Only spin up sector of the model is considered with $B=10$T.   a) Band structure of the system with periodic boundary conditions in all directions (no surfaces) projected onto the $z$ axis ($k$ denotes the crystal momentum along the $z$ direction). Solid (dashed) lines show occupied (empty) states. Occupation of the strained system is determined by adiabatically evolving the single-electron states of the unstrained system. b) Band structure of a slab with thickness $d=1000$\AA  \ (50 lattice sites). Only positive values of $k$ are displayed but the band structure is symmetric about $k=0$. Red (black) lines show occupied (empty) states. The central panel indicates the nonequilibrium occupancy of the strained  system obtained by adiabatically evolving the single-electron states of the unstrained system. The right panel shows the occupancy of the strained system once the electrons  relaxed back to equilibrium.  All three panels correspond to the same total number of electrons $N$. c) Change in the electron density in response to the applied strain as a function coordinate $y$ perpendicular to the slab surfaces. $\delta\rho$ refers to the nonequilibrium  distribution while $\delta\rho_{\rm eq}$ refers to the relaxed state.   Note that density oscillations near the edges apparent in $\delta\rho$ average to zero: there is no net charge transfer between the bulk and the surface in the nonequilibrium state, as can also be deduced from the vanishing $\delta\rho$ in the bulk.}
\label{fig8}
\end{figure*}
According to our previous discussion pseudoelectric field $\be$ should emerge when  the $u_{33}$ component of the strain tensor becomes time dependent. This can be achieved through dynamically stretching and compressing the crystal along its $z$ axis, e.g.\ by driving longitudinal sound waves through the crystal. To see how the lattice model realizes the chiral anomaly under these conditions we first consider an infinite bulk crystal in the presence of a uniform magnetic field $\bB=\hat{z}B$ and investigate the effect of the  static $u_{33}$ strain. The spectrum of an unstrained crystal in the field $B=10$T is displayed in Fig.\ \ref{fig8}a (we use once again Cd$_3$As$_2$ parameters and include this time also $\epsilon_\bk$). At low energies the spectrum exhibits the expected chiral branches that result from the $n=0$ Dirac Landau level. We assume the system is initially in its ground state with all energy levels below the chemical potential $\mu_0$ occupied and all levels above $\mu_0$ empty. We now implement unidirectional strain through Eq.\  (\ref{h5}) which amounts to rescaling the hopping amplitudes $t_1\to t_1(1-\alpha)$ and $c_1\to c_1(1-\alpha)$. {  Here $c_1$ is the hopping amplitude along the $z$ direction in $\epsilon_\bk$ defined below Eq.\ (\ref{ap1}).}  We imagine doing this sufficiently slowly so that the ground state evolves adiabatically in response to the increasing strain. The new ground state for $\alpha=0.03$ is depicted in  Fig.\ \ref{fig8}a. It exhibits a slightly modified band structure with the chemical potential shifted to a new value $\mu'$. The shift in $\mu$ occurs because under adiabatic evolution an electron initially in the quantum state with momentum $k$ in the $n$th band remains in that state as the band energy $E_n(k)$ evolves in response to strain. 

From the point of view of the low-energy theory the lateral shift of the chiral branches is consistent with the effect of the uniform chiral gauge potential $a_z$ which according to our discussion below Eq.\ (\ref{h4}) moves the Weyl points closer together for $\alpha>0$. From Eqs.\ (\ref{h6}) and (\ref{h7}) we can estimate the amount of this shift  $\delta Q\simeq (e/\hbar c)a_z=-u_{33}\cot{aQ}/a$. This in turn gives an estimate for the required change in the chemical potential $\delta\mu=\mu'-\mu_0=-\hbar v\delta Q$, or
\bee\label{p0}
\delta\mu=-{v\over c}e a_z=\alpha{\hbar v\over a}\cot{aQ}. 
\ee
For Cd$_3$As$_2$ parameters including the particle-hole symmetry breaking terms in $\epsilon_\bk$ we have $\hbar v\simeq 1.94$eV\AA\ which implies $\delta\mu=3.75$meV for $\alpha=0.03$. This estimate compares favorably with the value $\delta\mu_{\rm num}=3.46$meV obtained from our lattice model simulation presented in Fig.\ \ref{fig8}a.

If we continue focusing solely on the low energy degrees of freedom we would conclude that a change $\delta\mu$ in the chemical potential in a linearly dispersing band with degeneracy $(B/\Phi_0)$ brings about a change in the electron density 
\bee\label{p1}
\delta\rho=2{\delta\mu\over 2\pi \hbar v}\left({B \over\Phi_0}\right),
\ee
where the factor of 2 accounts for two chiral branches. 
Using Eq.\ (\ref{p0}) it is easy to verify that Eq.\ (\ref{p1}) coincides exactly with the prediction of the second chiral anomaly equation (\ref{an2}) for uniform static magnetic field and a time dependent pseudoelectric field $\be=-{1\over c}\partial_t\ba$. 

If on the other hand we espouse a band theory point of view then we see that in reality the charge density remains unchanged. This is because precisely the same number of single electron states are filled before and after the deformation. The chemical potential changes in order  to accommodate the fixed number of electrons in the modified band structure. We may thus conclude that in an infinite crystal pseudoelectric field induced by strain does not bring about any change in charge density. The chiral anomaly equation (\ref{an2}) however correctly predicts the strain induced change in the chemical potential $\delta\mu$. 

A change in the chemical potential, even if time dependent (as would be the case when strain is induced by a sound wave), is not easily measurable when not accompanied by a density change. So it would seem that this effect does not have observable consequences. Consider however a finite system with boundaries. The key point is that topologically protected surface states that are present in a Weyl semimetal will generally not respond to strain in the same way as the bulk states. To a good approximation one may consider the surface state to remain basically unaffected by a small  unidirectional strain. This is verified by our numerical simulations summarized in Fig.\ \ref{fig8}b. In that case application of strain will bring about a nonequilibrium distribution of electrons ($\mu$ changes in the bulk but remains unchanged at the surface). This is illustrated in Fig.\ \ref{fig8}b where we simulate the effect of a 3\% strain in a slab of thickness $d$ with surfaces perpendicular to the $y$ direction and magnetic field along $z$. We observe that strain shifts the chemical potential for the bulk states by the same amount as in the infinite system but leaves it essentially unchanged for the surface states.

Several interesting consequences follow from the above observation. First, we may expect the charge density to remain essentially unchanged in the strained crystal with nonequilibrium distribution of electrons. This is because the bulk density remains unchanged (as per our discussion above)  and since the total charge is conserved there can be no charge transfer to the surface. Second, in a real material the nonequilibrium electron distribution brought about by strain will relax towards equilibrium, causing dissipation in the system which is in principle observable. When the strain is induced by a sound wave this dissipation will provide a new mechanism for sound attenuation related to the chiral anomaly. Third, the relaxed charge density $\rho'(y)$  in the strained crystal will differ from the the original charge density $\rho_0(y)$ of the unstrained crystal because relaxation necessarily involves transfer of charge between the bulk and the surface of the sample. This is illustrated in Fig.\ \ref{fig8}c which shows the numerically calculated change in the charge density $\delta\rho(y)=\rho'(y)-\rho_0(y)$ in both nonequilibrium and equilibrium state following the application of a 3\% strain. We note that modulo some local fluctuations near the edge the charge density indeed behaves as expected on the basis of the above arguments.  

We conclude by elaborating on this last effect. If the sound frequency $\omega$ is small compared to the electron relaxation rate $\tau^{-1}$, as it will be the case in the typical experimental situation, the electron distribution will always remain close to  an equilibrium characterized by a global chemical potential $\mu'_{\rm eq}$.  The corresponding charge density should then exhibit significant variations as the chemical potential oscillates. 
Such a time dependent variation in the charge density will produce EM fields outside the sample which are measurable and can provide direct experimental evidence for the strain-induced chiral anomaly.  We shall estimate the distribution and the amplitude of these fields in the next Section.

To this end it will be useful to estimate the chemical potential $\mu'_{\rm eq}$ of the equilibrated strained system (see also  Fig.\ \ref{fig8}b). A straightforward calculation for a slab of thickness $d$ (summarized in Appendix B) gives
\bee\label{p2}
\mu'_{\rm eq}=\mu_0+{\delta\mu\over 1+\xi_B/d}, 
\ee
where $\xi_B=2Q\ell_B^2$ is the characteristic lengthscale and $\delta\mu$ is the chemical potential change in the system without surfaces given by Eq.\ (\ref{p0}). The physics of  Eq.\ (\ref{p2}) is quite simple: it reflects the fact that a surface can accommodate only a limited amount of charge from the bulk. For a thick slab $d\gg \xi_B$ we recover the bulk result $\mu'_{\rm eq}\approx \mu_0+\delta\mu$ because the effect of the surface becomes negligible.

 From Eq.\ (\ref{p2}) it is easy to obtain an estimate for the corresponding  change in the bulk charge density
\bee\label{p3}
\delta\rho^{\rm bulk}=-{\alpha\over \pi a}\left({B\over \Phi_0}\right){\cot{aQ}\over 1+d/\xi_B}.
\ee
In the limit of a thin slab, $d\ll \xi_B$, this result approaches the charge density change (\ref{p1}) derived based on the naive application of the chiral anomaly equation, except for the opposite overall sign. In this limit, physically, almost all the non-equilibrium charge density produced in the bulk can be absorbed by the surface. The bulk charge density thus goes down by the amount close to that predicted by the chiral anomaly.   

\begin{figure}[t]
\includegraphics[width = 7.5cm]{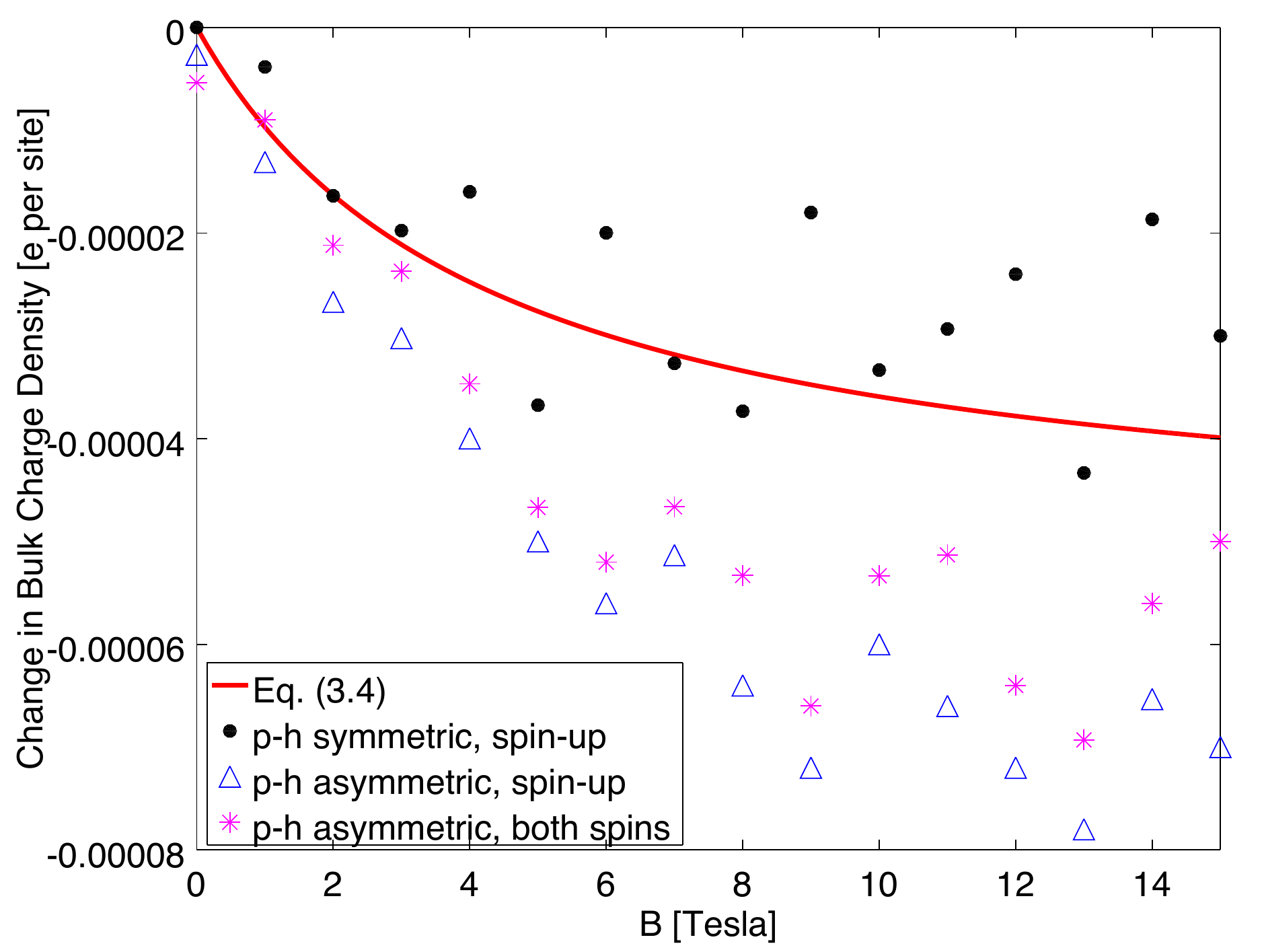}
\caption{Numerically calculated change in the bulk charge density $\delta\rho^{\rm bulk}$ in response to unidirectional strain $\alpha=0.03$ as a function of the applied field $B$. Parameters for Cd$_3$As$_2$ are used with $\mu=0$ and $d=1000$\AA  \ (50 lattice sites). Solid black symbols give result for the p-h symmetric version of the ${1\over 2}$-Cd$_3$As$_2$ model obtained by setting all $C_j$ parameters to zero.
}\label{fig10}
\end{figure}
Fig.\ \ref{fig10} shows the bulk charge density $\delta\rho^{\rm bulk}$ in response to the unidirectional strain $\alpha=0.03$ as a function of the applied field $B$ in a relaxed system, numerically calculated from the lattice model. A good agreement with Eq.\ (\ref{p3}) is seen both in the magnitude of the effect and  its functional form. The lattice model shows a somewhat stronger response than expected which we attribute to the p-h anisotropy that was not included in the analytical calculation. That this is indeed the case is confirmed by the same calculation performed for  the p-h symmetric version of the ${1\over 2}$-Cd$_3$As$_2$ model which shows closer agreement with Eq.\ (\ref{p3}), modulo finite size effect induced fluctuations (solid black symbols in Fig.\ \ref{fig10}). We however note that in this case the contribution from the spin-down sector exactly cancels that from spin up so p-h asymmetry is required to obtain a nonzero result.

We note that Eqs.\ (\ref{p2}) and (\ref{p3}) were derived assuming quantum limit, i.e\ chemical potential in the $n=0$ Landau level. The corresponding results valid away from the quantum limit are given in Appendix B.

\begin{figure}[t]
\includegraphics[width = 5.5cm]{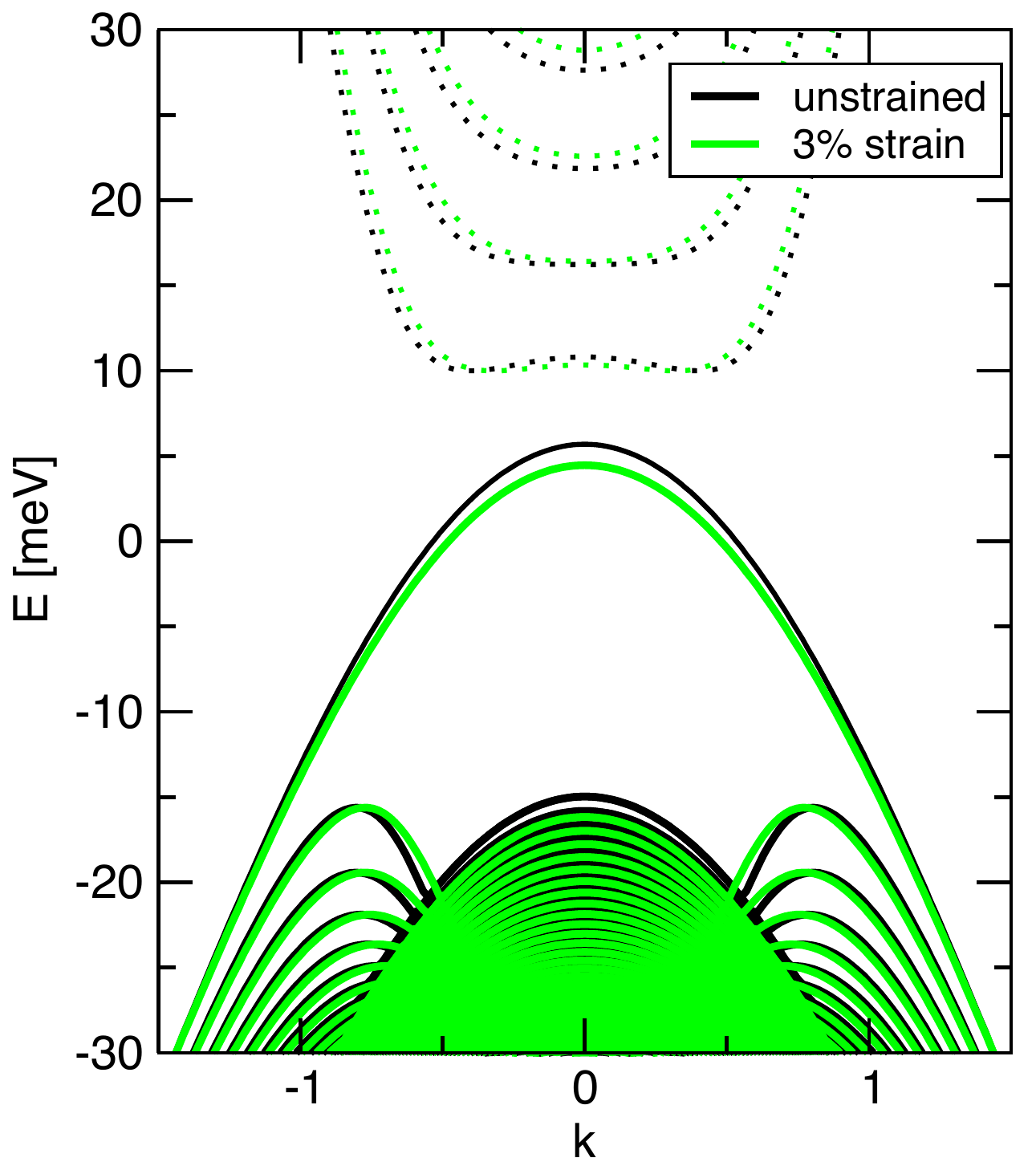}
\caption{Band structure of the spin down sector of Cd$_3$As$_2$ in magnetic field $B=10$T. Two chiral branches are visible at low energy but they are now strongly distorted by p-h symmetry breaking terms and they no longer traverse the gap between the valence and the conduction band.
}\label{fig9}
\end{figure}
We close this subsection by considering Dirac semimetals. Naively one could think that the effects discussed above will cancel once we include both spin sectors.  This would indeed be the case in a perfectly particle-hole symmetric system. However, the band structures of both Cd$_3$As$_2$ and Na$_3$Bi exhibit strong particle-hole asymmetry which prevents such cancellations. To elucidate this we show in Fig.\ \ref{fig9} the band structure of the spin-down sector of Cd$_3$As$_2$ in the field of 10T. Compared to the spin-up sector (Fig.\ \ref{fig8}a) it indicates a spectral gap at low energies. It is clear that when $\mu$ lies inside this gap then all the physics comes exclusively from the spin-up sector. Specifically, there is nothing here to cancel or even weaken the effects discussed above. We find that this remains true more generally. Even when $\mu$ is outside the gap the contributions to various effects discussed above generically do not cancel but remain of a similar magnitude as they would be in a Weyl semimetal with a single pair of Weyl points. This is illustrated in Fig.\ \ref{fig10} where the chemical potential is chosen to lie outside the bandgap; the effect is only slightly reduced when contributions from both sectors are added up. We thus expect the effects discussed above to generically remain present in Dirac semimetals such as Cd$_3$As$_2$ and Na$_3$Bi.

%%%%%
\section{Experimental manifestations of the strain-induced chiral anomaly}
%

%%%
\subsection{Persistent currents in a twisted Weyl semimetal with broken $\cT$: topological coaxial cable} 

\begin{figure}[t]
\includegraphics[width = 7.5cm]{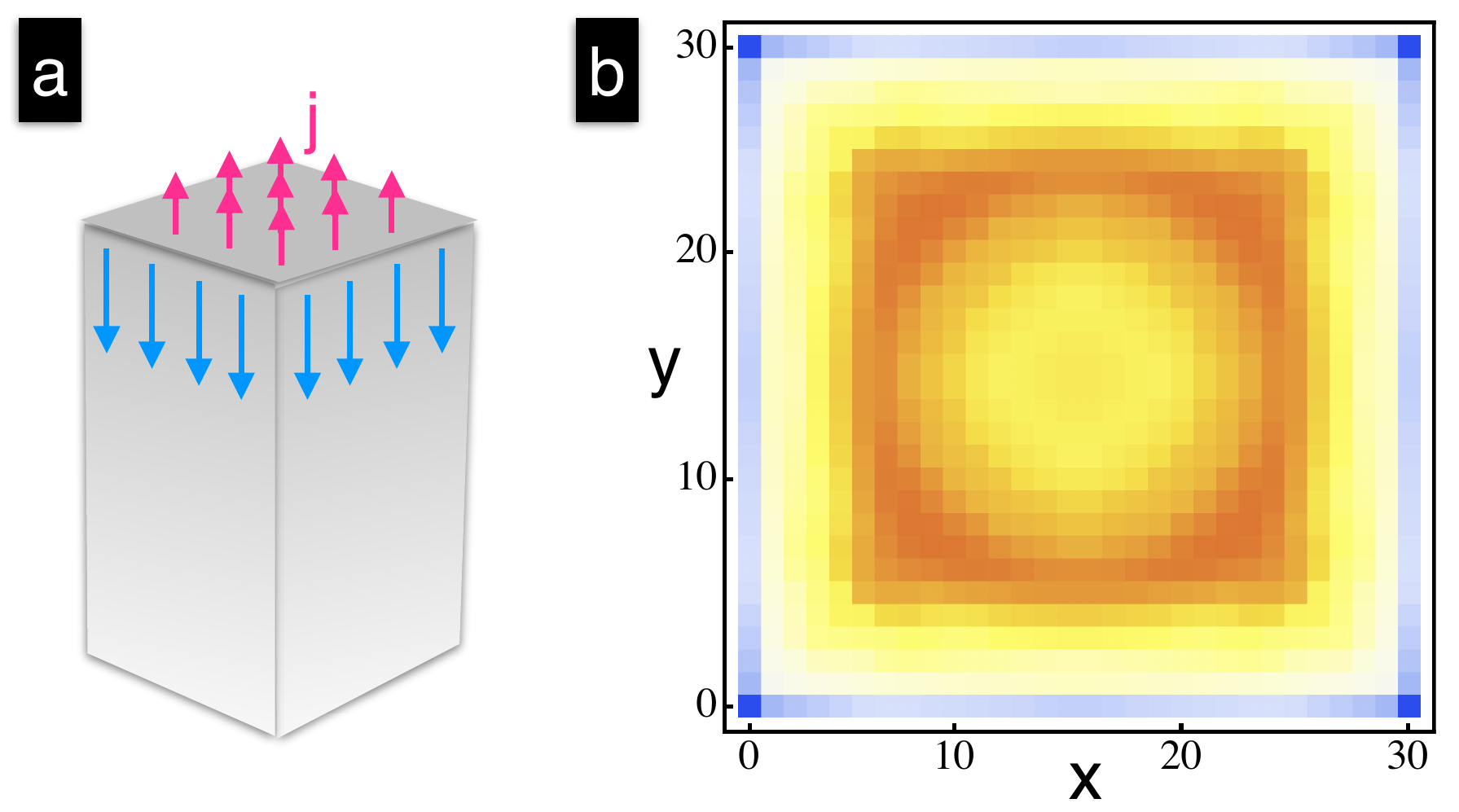}
\caption{Equilibrium current density in the Weyl semimetal wire under torsion. a) Schematic depiction of the bulk/surface current flow. b) Ground state current density computed from the lattice model Eqs.\ (\ref{h3})  and (\ref{h5}) at chemical potential $\mu=5$meV. Warm (cold)  colors represent positive (negative) current density $\bj$.
{  The ring-shaped inhomogeneity in $\bj$  apparent in the bulk of the wire reflects Friedel-like oscillations in electron wavefunctions caused by the presence of the surface.} 
}\label{fig3}
\end{figure}

The phenomena discussed above have several observable consequences which we now discuss. According to Fig.\ \ref{fig2}c  Weyl semimetal wire under torsion exhibits spatial separation between left and right moving modes at low energies: the former are localized near the boundary while the latter occur in the bulk. At a generic chemical potential we thus expect persistent equilibrium currents to flow in such a wire as indicated in Fig.\ \ref{fig3}a. This can be argued as follows. Suppose the current density $j_z(\br)$ is uniformly zero at some reference chemical potential $\mu_0$. If we now change the chemical potential to $\mu=\mu_0+\delta\mu$ we are populating additional right moving modes in the bulk and left moving modes at the surface of the wire.   Although the total current carried by the wire remains zero, as it must be in any normal metal in equilibrium \cite{vazifeh2013},  there is now a non-vanishing positive current density flowing in the bulk compensated by the negative current density flowing along the surface. We have verified numerically that this is indeed the case in the lattice model  (\ref{h3})  and (\ref{h5}): for any chemical potential  $\mu\neq 0$ a ground-state current density develops as illustrated in Fig.\ \ref{fig3}b.    

Such a current flow generates magnetic fields outside the wire which are, at least in principle, measurable e.g\ by scanning SQUID microscopy. In practice, however, we expect this to be a challenging experiment. The currents occur only in a Weyl semimetal with broken $\cT$ which is most likely to be realized in a magnetic material. It might be difficult to distinguish the  fields produced by torsion-induced persistent currents from the sample magnetization. We note that in Dirac semimetals, like Cd$_3$As$_2$ or Na$_3$Bi, the total current density will vanish upon including the contribution from the lower diagonal block in the Hamiltonian (\ref{h2}). This has to be the case because non-zero $\bj$ would violate the $\cT$ symmetry of the material, which should remain unbroken under strain. The current density can be nonzero, however, when both torsion and magnetic field are applied. This is demonstrated in Fig.\ \ref{fig3_supp} of Appendix A.

%%%
\subsection{Chiral torsional effect}

The physics described above however has a simple manifestation observable in  transport measurements in both Weyl and Dirac semimetals. Consider a measurement of longitudinal resistivity in a twisted wire. Once again we start by discussing a Weyl semimetal. When electric field $\bE$ is applied to the twisted wire it begins to produce  charge density $\delta\rho=\rho-\rho_0$ in the bulk at the rate given by the anomaly equation   (\ref{an2}). In view of our discussion above we interpret $\delta\rho$ as charge density imbalance between the bulk and the surface of the wire. Such an imbalance can relax back to equilibrium only through processes that induce backscattering between the bulk right moving modes and the surface left moving modes. If we denote the relevant scattering time by $\tau$ we get an equation 
\bee \label{h11}
{d\over dt}\delta\rho={e^2\over 2\pi^2\hbar^2c}\bE\cdot\bb-{\delta\rho\over\tau}. 
\ee
At long times $t\gg\tau$ the steady state solution reads
\bee \label{h12}
\delta\rho={e^2\tau\over 2\pi^2\hbar^2c}\bE\cdot\bb. 
\ee
The wire clearly carries non-zero electrical current.
The expression for the current depends on the relative position of the chemical potential $\mu$ and the bottom of the first Landau level $\epsilon_1(0)=\hbar v\sqrt{2eb/\hbar c}$. In the quantum limit, $|\mu|< \epsilon_1(0)$, only the chiral modes in the $n=0$ Landau level are populated. These all move at the same velocity $v\sgn(b)$ and the non-equilibrium charge density $\delta\rho$ thus gives electrical current 
\bee\label{h13}
J_{\rm CTE}=-ev\sgn(b)\delta\rho={e^3v\tau\over 2\pi^2\hbar^2c}\bE\cdot \bb\sgn(b).
\ee

For a constant relaxation time $\tau$ we thus have chiral torsional contribution to the conductivity $\sigma_{\rm CTE}\sim |b|$, similar to the ordinary chiral magnetic effect  $\sigma_{\rm CME}\sim |B|$ in the quantum limit \cite{nielsen1983}. However, if the wire radius $R$ significantly exceeds the magnetic length $\ell_b=\sqrt{\hbar c/eb}{  \ \simeq 256{\rm \AA}\sqrt{1{\rm T}/b}}$, then we find that the appropriate relaxation time becomes field-dependent, namely 
\bee\label{h14}
\tau\simeq \tau_0{R^2\over\ell_b^2}\sim |b|,
\ee
where $\tau_0$ is the microscopic scattering time.
This is because the bulk electron wavefunctions have spatial extent $\ell_b$ in the direction transverse to the axis of the wire. Deep in the bulk impurities cause scattering between the individual bulk modes but since these are all right moving such processes cannot relax the current. Only those electrons that have diffused all the way to the boundary through repeated scattering processes can backscatter into left moving surface modes. Electrons thus experience hydrodynamic flow whereby dissipation occurs only at the boundary.  Eq.\ (\ref{h14}) is derived in Appendix C and expresses the fact that an electron that is produced near the center of the wire has to travel distance $R$ to the boundary and this takes on average $(R/\ell_b)^2$ scattering events. We conclude that 
$\sigma_{\rm CTE}\sim b^2$ in the quantum limit when $\ell_b\ll R$.

In the semiclassical limit, $|\mu|\gg \epsilon_1(0)$, we must take into account the additional equilibration that occurs between the individual Landau levels within a given Weyl point. We assume that the relaxation time for this process is very short and essentially instantaneous compared to $\tau$. In this case, electron density produced through Eq.\ (\ref{h12}) gets distributed among all the bulk states and leads to a shift in the chemical potential $\mu\to \mu+\delta\mu$. In the semiclassical limit we can approximate the density of states by the expression valid in the zero field, $D(\epsilon)=\epsilon^2/\pi^2\hbar^3v^3$, where for simplicity we also assume isotropic velocities. In the limit of interest, $\delta\mu\ll k_BT\ll\mu$, it is easy to find from this the shift in the chemical potential caused by a small change in density,
\bee\label{h15}
\delta\mu\simeq{2\pi^2\hbar^3v^3\over \mu^2+{2\pi^2\over 3}k_B^2T^2}\delta\rho.
\ee
We can now calculate the current by noting that, once again, only the chiral branches contribute. We thus obtain 
\bee\label{h16}
J_{\rm CTE}=-ev\left({\delta\mu\over 2\pi\hbar v}\right)\left({b\over\Phi_0}\right),
\ee
where the first bracket represents the number of extra modes $\delta n$ that have been populated on the chiral branch and the second reflects their degeneracy. Combining this with Eqs.\ (\ref{h15}) and  (\ref{h12}) we find
\bee\label{h17}
J_{\rm CTE}={e^4v^3\over 8\pi^3\hbar c^2}{\tau\over \mu^2+{2\pi^2\over 3}k_B^2T^2}(\bE\cdot\bb)b.
\ee

In view of Eq.\ (\ref{h14}) in a Weyl semimetal under torsion (parametrized here by $b\propto \Omega$) we thus predict a positive contribution to the conductivity 
\bee\label{h18}
\sigma_{\rm CTE}\propto\left\{
\begin{array}{lll}
b^2, &\mu<\hbar v\sqrt{2eb\over\hbar c} \  & {\rm quantum \ limit}, \\
|b|^3,  &\mu\gg\hbar v\sqrt{2eb\over\hbar c} \ & {\rm semiclassical \ limit}.
\end{array}
\right.
\ee
The predicted field dependence  is different from the analogous effect encountered in the presence of the real magnetic field $B$  (where $\sigma_{\rm CME}$ behaves as $\sim B$ and $\sim B^2$ in the two limits). This reflects the hydrodynamic nature of the electron flow that occurs when  $R\gg \ell_b$.  The right and left moving modes are then segregated to the bulk and the boundary respectively, which leads to an extra power of $b$ due to $b$-dependent transport scattering rate (\ref{h14}). We also note that when  $R\gg \ell_b$, Eq.\  (\ref{h14}) implies significant enhancement of the transport lifetime and thus leads us to expect a strong effect. In the opposite limit, $R\lesssim \ell_b$, the transport scattering rate becomes field independent and the more conventional behavior with $\sigma_{\rm CTE}\propto b\  (b^2)$ in quantum (semiclassical) limit is restored.

The effect will persist in a Dirac semimetal such as Cd$_3$As$_2$ and Na$_3$Bi, which can be thought of as two $\cT$-conjugate copies of the Weyl semimetal discussed above. In the presence of a twist the spectrum will consist of that indicated in Fig.\ \ref{fig2}c for the spin-up sector plus a time-reversed copy (obtained by reversing $k\to -k$) for the spin down sector. The same analysis we just performed applies unchanged for each spin sector if one can ignore spin-flip scattering events. In this case Eq.\ (\ref{h18}) continues to hold in a Dirac semimetal. Spin-flip processes, if present, open additional channel for relaxation by scattering between left and right moving bulk modes. In the limit when the spin-flip relaxation rate $\tau_{\rm sf}^{-1}$ exceeds $\tau^{-1}$ the hydrodynamic flow will cease and the behavior will cross over to the regular chiral anomaly with $\sigma_{\rm CTE}\propto b\ (b^2)$ in the quantum (semiclassical) limit.  In clean samples of $\cT$-preserving Cd$_3$As$_2$ and Na$_3$Bi we expect the hydrodynamic behavior to prevail. This is because ordinary non-magnetic impurities cannot cause spin-flip scattering. Time reversal symmetry permits spin-orbit scattering  terms of the form $\hat{z}\cdot(\bsig\times\bk)$. These do contribute to $\tau_{\rm sf}^{-1}$  but we expect such contributions to be small.

%%%
\subsection{Ultrasonic attenuation and  EM field emission}

We now consider the experimental manifestations of the $\be\cdot\bB$ term in the second chiral anomaly equation (\ref{an2}). For concreteness we again start with a Weyl semimetal and consider a sample in the shape of a  slab with thickness $d$ and surfaces perpendicular to the $y$ axis. Magnetic field $B$ is applied  along the $z$-direction. The requisite $\be$ field is generated by a longitudinal sound wave with frequency $\omega$ that is driven along the $z$ direction. This produces a time dependent displacement field 
\bee\label{h20}
\bu=u_0\hat{z}\sin{(qz-\omega t)},
\ee
where $q=\omega/c_s$ is the wavenumber and $c_s$ the sound velocity. The nonzero component of the strain tensor is $u_{33}=u_0q\cos{(qz-\omega t)}$ which through Eq.\ 
(\ref{p0}) yields an oscillating component of the bulk chemical potential
\bee\label{j21}
\delta\mu(t)=u_0q\left({\hbar v\over a}\cot{aQ}\right)\cos{(qz-\omega t)}. 
\ee

As mentioned in Sec.\ III.B electron relaxation dissipates energy which will be manifested by the  attenuation of the sound wave as it propagates through the medium. Specifically, as explained e.g.\ in Ref.\ \cite{abrikosov} the energy flux $I$ carried by the sound wave obeys $I(z)=I_0 e^{-2\Gamma z}$ where $\Gamma$ is the sound attenuation coefficient.  We now proceed to estimate $\Gamma$ which is given by $\Gamma=Q/2I$, where $Q$ denotes the amount of energy dissipated in a unit volume per unit time. To provide a crude estimate of $Q$ we assume for a moment that 
the electron relaxation rate $\tau^{-1}$ is comparable to the driving frequency, $\omega\tau\approx 1$. In this case relaxational dynamics is maximally out of phase with the sound wave and we can estimate $Q$ simply by calculating the energy difference between the nonequilibrium distribution of electrons (see Fig.\ \ref{fig8}b) reached at the crest of the wave (assuming no dissipation has occurred until then) and the corresponding equilibrium distribution with the chemical potential $\mu'_{\rm eq}$. For this estimate  consider a slice of the system perpendicular to $z$ of length $l$ such that $l\ll\lambda_s$. Inside the slice the strain can be considered uniform, implying a
uniform chemical potential $\delta\mu(t)\propto\cos{\omega t}$.
We may thus estimate the total dissipated electron energy per cycle as 
\bee\label{j22}
E_{\rm dis}= lwd\int_{\mu'_{\rm eq}}^{\mu'}\epsilon D_b(\epsilon) d\epsilon -
lw\int_{\mu_{0}}^{\mu'_{\rm eq}} \epsilon D_s(\epsilon) d\epsilon,
\ee
where $w$ is the width of the slab along the $x$ direction and $D_{b/s}(\epsilon)$ is the bulk/surface density of states given in Appendix B.  It is easy to evaluate the requisite integrals. After some algebra and with help of Eq.\ (\ref{p2}) one obtains, assuming quantum limit,
\bee\label{j23}
E_{\rm dis}\approx {lwd\over2\pi \hbar v}\left({B\over \Phi_0}\right) {1\over 1+d/\xi_B}\delta\bar{\mu}^2, 
\ee
where $\delta\bar{\mu}$ is the amplitude of $\delta\mu(t)$ given in Eq.\ (\ref{j21}).

A more complete treatment of the relaxational dynamics, which we omit here for the sake of brevity, gives a result for the energy dissipated per cycle valid for any frequency
\bee\label{j24}
E_{\rm dis}={lwd\over2\pi \hbar v}\left({B\over \Phi_0}\right) {\omega\tau\over (1+d/\xi_B)^2+(\omega\tau)^2}\delta\bar{\mu}^2.
\ee

The energy density of the sound wave, averaged over one cycle, is $\rho_E={1\over 2}\rho c_s^2u_0^2q^2$, where $\rho$ denotes the mass density of the crystal. Noting that the corresponding energy flux is $I=c_s\rho_E$ one obtains the sound attenuation coefficient
\bee\label{j25}
\Gamma={(\omega E_{\rm dis}/lwd)\over 2 c_s\rho_E}.
\ee
To estimate its magnitude we assume the limit of a thin slab $d\ll\xi_B$ and fast relaxation $\omega\tau\ll 1$ in Eq.\ (\ref{j24}).  In this limit $\Gamma$ becomes independent of $d$:
\bee\label{j26}
\Gamma\simeq \left({\omega\over 2\pi c_s}\right){\hbar v\over a^2} 
\left({B\over \Phi_0}\right){2\cot^2{aQ}\over \rho c_s^2}(\omega\tau).
\ee
For our estimate we take $f=\omega/2\pi=200$MHz, the mass density of Cd$_3$As$_2$ is $\rho=7.0\times 10^3$ kg/m$^3$ while the speed of sound is $c_s=2.3\times 10^3$m/s \cite{2007MAW200717} which gives $\lambda_s\simeq 11\mu$m at this frequency. For these parameters we obtain
\bee\label{j27}
\Gamma\simeq 3.6\times 10^3 {\rm m}^{-1}\left[{B\over 1{\rm T}}\right](\omega\tau).
\ee

We see that depending on the magnitude of the electron scattering rate the sound attenuation can be substantial.   There are of course many conventional  sources of ultrasonic attenuation in metals \cite{abrikosov}. Given the specific dependence of $\Gamma$ on frequency, magnetic field and the fact that it depends only on the component of  $\bB$ parallel to $\bq$,  it should be possible to separate the contribution of the chiral anomaly from the  more conventional contributions. 

At $B=1$T for material parameters relevant to Cd$_3$As$_2$ we have $\xi_B\simeq 430$nm so the above estimate applies to thin films or wires. For thicker films  one must include the additional suppression factor $(1+d/\xi_B)^{-2}$ from Eq. (\ref{j24}) that we neglected so far. This factor reflects the fact that the relaxation mechanism involves charge transfer from the bulk to the surface of the sample. For the same reason, however, we expect in this limit to obtain an enhanced relaxation time $\tau\simeq\tau_0(d/\ell_B)^2$, where $\tau_0$ is the microscopic relaxation time as in Eq.\ (\ref{h14}). This is because to relax the non-equilibrium distribution brought about by the chiral anomaly bulk electrons must diffuse to the surface and this takes on average $(d/\ell_B)^2$ scattering events. In the end we expect only a weak dependence of $\Gamma$ on the sample thickness $d$ although a detailed treatment of the combined spatial and temporal distribution of electrons during the relaxation process is an interesting topic for future research.

The oscillating charge density that occurs in the system in response to the sound wave will generate EM fields that can be detected outside the sample. We show below that in a typical situation the electric field close to the surface can be substantial and thus detectable. The field decays as $\sim e^{-r/\lambda_s}$ away from the surface but since $\lambda_s$ is tens or hundreds of microns at typical ultrasonic  frequencies the detection of such fields should not be difficult \cite{note1}.

\begin{figure}[t]
\includegraphics[width = 7.5cm]{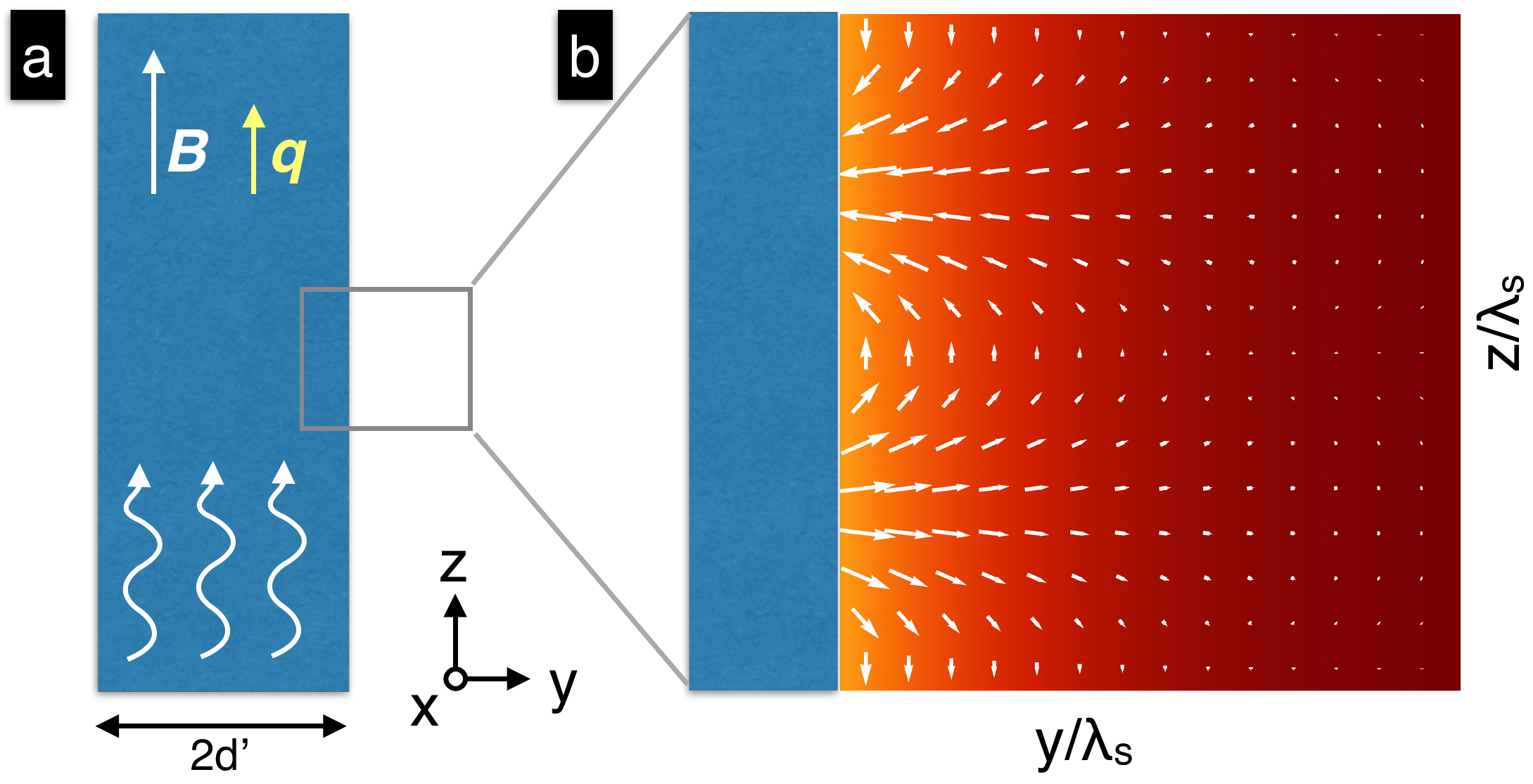}
\caption{Proposed geometry for the EM field emission measurement in the limit when all the dimensions of the crystal are much larger than the sound wavelength $\lambda_s$. a)  A slab of thickness $d=2d'$ is subjected to magnetic field $\bB$ and a longitudinal acoustic sound wave propagating along the $z$ direction. b) A snapshot of the electric field distribution near the surface calculated from Eq.\ (\ref{h30}).  As a function of time the entire pattern moves in the $z$ direction at the speed of sound $c_s$. 
}\label{fig6}
\end{figure}

To estimate the amplitude of the electric field we assume once again that electron relaxation is fast compared to the driving frequency, $\omega\tau\ll 1$. This means that electrons will locally always be close to equilibrium characterized by the charge density $\bar\rho+\delta\rho^{\rm bulk}(z,t)$ where $\bar\rho$ is the bulk charge density of the unstrained crystal and $\delta\rho^{\rm bulk}(z,t)$ is given by Eq.\ (\ref{p3}) with $\alpha$ now describing the local strain field at $(z,t)$.   In a slab of thickness $d=2d'$ illustrated in Fig.\ (\ref{fig6})  the oscillating component of the charge density therefore reads
\bee\label{h22}
\rho^{\rm bulk}=\rho_0\cos{(qz-\omega t)},
\ee
with 
\bee\label{h23}
\rho_0=-{u_0q\over \pi a}\left({B\over \Phi_0}\right) {\cot{aQ}\over (1+d/\xi_B)}.
\ee

From our previous discussion we know that the charge generated in the bulk comes from the boundary. The total charge density  that reflects the overall charge conservation in each constant-$z$  slice of the sample  can thus be written as
\bee\label{h27}
\rho=\rho_0\left[\theta(d'-|y|)-d'\delta(y\pm d')\right]\cos(qz-\omega t),
\ee
where $\rho_0$ is given by Eq.\ (\ref{h23}). In the near field (static) region we may neglect the magnetic effects and determine the electric field $\bE=-\nabla\Phi$ by solving the Poisson equation $\nabla^2\Phi=-4\pi\rho$.  Adopting the ansatz $\Phi(\br,t)=f(y)\cos(qz-\omega t)$ we are led to a 1D equation for $f(y)$ of the form 
\bee\label{h28}
(\partial^2_y-q^2)f=-4\pi\rho_0\left[\theta(d'-|y|)-d'\delta(y\pm d')\right].
\ee
This has a solution
\bee\label{h29}
f(y)=\left\{
\begin{array}{lll}
{4\pi\rho_0\over q^2}+B\cosh{qy}, \  &|y|<d', \\
Ae^{-q|y|},  & |y|>d'.
\end{array}
\right.
\ee
The function $f(y)$ must be continuous at $y=\pm d'$ and the discontinuity in its first derivative must match the surface charge in Eq.\ (\ref{h28}), $f'(d'+\epsilon)-f'(d'-\epsilon)=8\pi\rho_0 d'$. This determines  constants $A$ and $B$.  The full solution for the potential outside the sample reads 
\bee\label{h30}
\Phi(\br,t)={4\pi\rho_0} {Ae^{-q|y|}\over q^2}\cos(qz-\omega t),
\ee
with $A=\sinh{qd'}-2qd\cosh{qd'}\approx -qd'\  e^{qd'}$. The electric field that follows from this potential is depicted in Fig.\ \ref{fig6}b. For $d=1$mm, $u_0=0.01a$ and all the other parameters as before the maximum electric field (that occurs right at the sample surface) can be estimated as $|\bE|\simeq 4\pi\rho_0 ed\simeq 2.4\times 10^4$V/m.  This is a large field which should be easily detectable. 

In a realistic semimetal we should include screening effects  which can substantially reduce the electric field amplitude estimated above. A crude estimate of the screened field can be obtained by replacing $\Phi(q)\to\Phi(q)/\epsilon(q)$ where $\epsilon(q) =1+k_{\rm TF}^2/q^2$ is the dielectric function in the Thomas-Fermi approximation and $k_{\rm TF}^2= 4\pi e^2 D(\mu)$. It is physically more transparent to write $(k_{\rm TF}/q)^2=(\lambda_s/\lambda_{\rm TF})^2$ where $\lambda_{\rm TF}=1/k_{\rm TF}$ is the Thomas-Fermi screening length.  Using the experimentally determined electron velocity $v\simeq 1.5\times 10^6$m/s  \cite{neupane2014} to obtain density of states $D(\epsilon)=\epsilon^2/\pi^2\hbar^3 v^3$ we find that $\lambda_{\rm TF}\simeq 32 \mu{\rm m} [1 {\rm meV}/\mu]$. Thus, depending on the chemical potential, the screening length can be quite long. For instance if $\mu=10$meV the screening length $\lambda_{\rm TF}\simeq 3.2\mu$m is comparable to $\lambda_s=11\mu$m and the electric potential will be suppressed only modestly. Even for the experimentally observed $\mu\approx 200$meV \cite{neupane2014} the suppression is about a factor of $1.6 \times 10^4$ which still leaves a significant field strength  of several V/m at the surface. We conclude that the effect remains measurable even in the presence of realistic levels of screening that can be expected in a  Dirac semimetal with the chemical potential not too far from the neutrality point.

\subsection{$\be\cdot\bb$ term and the chiral anomaly in the absence of EM fields}

We finally mention an attractive possibility of  testing the chiral anomaly using purely strain-induced gauge fields and no real EM fields. According to our previous discussion a  simultaneous application of torsion and time-dependent unidirectional strain in a wire geometry generates both $\bb$ and $\be$ pointed along the $z$ direction of the crystal. In this situation the right hand side of the first anomaly equation (\ref{an1}) becomes nonzero even in the complete absence of $\bE$ and $\bB$ and pumping of charge between the Weyl nodes occurs. The nonequilibrium electron distribution thus created will relax via internodal scattering and produce dissipation of energy. This dissipation is in principle measurable. For instance when the pseudoelectric field $\be$ is generated by a longitudinal sound wave its amplitude will be attenuated by this effect and the attenuation coefficient will be proportional to the amount of torsion on the wire. This can be demonstrated by considerations that are similar to those that lead to Eq.\ (\ref{j26}). We shall not repeat this analysis here but simply note that a substantial  contribution to the attenuation can be achieved by this effect.

\section{Conclusions and outlook}

We studied the chiral anomaly in Weyl semimetals in a new context, when the sign of the anomaly is the same in the two Weyl cones. This takes place when a chiral gauge field is present in addition to the ordinary EM gauge field. Specifically, this type of chiral anomaly  occurs when pseudomagnetic field $\bb$, produced by torsion in the material, is present together with the real electric field $\bE$. Alternatively, pseudoelectric $\be$ field produced by unidirectional strain combined with a real magnetic field $\bB$ gives rise to the anomaly. 
Contrary to the usually discussed chiral anomaly, density of electron grows in \textit{both} Weyl cones when the fields are applied, thus making the bulk theory of the material truly anomalous. The apparent contradiction with charge conservation is resolved when one takes into account the surface of the material -- fermions are taken from the surface into the bulk.

In the presence of the $\bb\cdot\bE$ term the transfer of charge from the surface to the bulk occurs through the ordinary semiclassical evolution of the electron states in the Brillouin zone. This is facilitated by the phenomenon of spatial segregation of the right and left moving modes between the surface and the bulk of the wire under torsion as discussed in Sec. III.A.  In the presence of the $\be\cdot\bB$ term the situation is different: here the charge transfer occurs through relaxation of the nonequilibrium state that is generated by the chiral anomaly. This disparity in the action of the two types of terms has a very simple physical reason. A uniform $\bb$ field can only exist in a finite system with boundaries because it requires an increasing strain field in some spatial directions (just like the uniform $\bB$ field requires increasing vector potential $\bA$). However, in a realistic crystal strain can only increase to a certain point after which the crystal breaks. We thus see that a uniform $\bb$ necessarily implies the existence of surfaces. A consequence of this is that the band structure of the relevant system will have an equal number of left and right moving modes which are however unbalanced between the surface and the bulk. Semiclassical evolution in the presence of $\bE\parallel\bb$ will thus transfer charge from the bulk to the surface (or vice versa). By contrast the $\be$ field can be created by a time-dependent unidirectional strain which does not require spatial boundaries. We have seen that in a system without boundaries the $\be\cdot\bB$ term simply changes the chemical potential in accord with the chiral anomaly equation (\ref{an2}). If, however surfaces exist, this creates a nonequilibrium distribution which can relax by transferring charge to the surface. Furthermore, if sufficient number of surface states are available, then the bulk charge density change can be close to that predicted by the chiral anomaly.      

When both torsion and unidirectional strain are applied a new form of the chiral anomaly can be created via the $\be\cdot\bb$ term in Eq.\ (\ref{an1}). In this case charge is transferred between the two Weyl points with opposite chirality but remarkably no physical  EM fields are required.

Based on these general concepts we make several specific predictions for the experimentally observable signatures of the anomaly. In the case of the $\bb\cdot\bE$ term we predict a negative contribution to the resistance  that has a square or cubic dependence on the torsion strength, depending on the regime. In the case of the $\be\cdot\bB$ term we predict bulk-boundary charge transfer, resulting in EM field emission and ultrasonic attenuation. Similarly, we predict that the $\be\cdot\bb$ term will contribute to ultrasonic attenuation. These predictions are most clear cut in a semimetal with a single pair of Weyl points. We showed, however, that substantial observable consequences occur also in Dirac semimetals Cd$_3$As$_2$ and Na$_3$Bi whose electronic structure can be viewed as two time-reversed copies of such an elemental Weyl semimetal. On general grounds we also expect these phenomena to be manifested in more complex Weyl semimetals such as TaAs, ZrTe$_5$ or WTe$_2$ which host several pairs of Weyl points. 
{  Because their electronic structures are complex, detailed  quantitative modeling of strain effects will require delving into the details of the band structures and we leave this for future study. We nevertheless
anticipate that in these materials  each pair of Weyl points will contribute to various chiral anomaly related effects discussed in this paper. The contributions will have different magnitudes and signs depending on the relative positions and Fermi velocities of the Weyl cones. Partial cancellations can occur but it appears unlikely that the effect would vanish completely, except perhaps for very specific strain patterns with high symmetry. Pronounced transport signatures of the ordinary chiral anomaly have already been detected in several materials \cite{kim2013,huang2015,ong2015,valla2016,jia2016} including some with multiple Weyl points. This strongly suggests that the novel strain-induced effects predicted in this work should also be observable in these materials.}

Our work draws upon several previous studies. Some of our results regarding the physical consequences of the second anomaly equation (\ref{an2}) have been foreshadowed in Ref.\ \cite{qi2013} which considered Weyl fermions in magnetically doped topological insulators. Here the chiral gauge field can arise from magnetic fluctuations in the system and was predicted to produce one-dimensional chiral modes in a ferromagnetic vortex line and a novel plasmon-magnon coupling. As far as we know Weyl fermions have not yet been observed in magnetically doped topological insulators. Also, it may be challenging to create and control the magnetic textures envisioned in Ref.\ \cite{qi2013}. By contrast the phenomena predicted in our work only require existing materials, such as Cd$_3$As$_2$ or Na$_3$Bi. Also, producing the chiral gauge field from strain is not expected to pose an exceptional experimental challenge.  Our work also draws upon the results of Refs.\ \cite{guinea2010,cortijo2015,shapourian2015} which established the equivalence between strain and chiral gauge field in various materials ranging from graphene and topological insulators to a simple model of a Weyl semimetal. Our study however goes far beyond the scope of Ref.\ \cite{cortijo2015} by considering the effect of strain in specific materials and geometries and by providing concrete quantitative predictions for experimentally measurable quantities related to the chiral anomaly. 

When our work was substantially completed we became aware of a preprint \cite{fujimoto2016} which discussed a fictitious magnetic field in a Weyl semimetal created by crystal dislocations. This effect is closely related to our $\bb$ field but is different in that unlike externally applied strain, dislocations in a crystal cannot be easily controlled. Therefore, experimental detection of this effect may prove challenging. Very recently Schuster {\em et al.} \cite{chamon2016} discussed the concept of a topological coaxial cable in a gapped Weyl semimetal with a vortex in the Higgs field that is responsible for the gap. In this situation the vortex line is predicted to carry protected fermionic modes and contribute exactly quantized conductance. This effect is very interesting but also very different from our concept of topological coaxial cable which occurs in an ungapped Dirac or Weyl semimetal and does not in general produce quantized conductance.

Given both the fundamental nature of the new anomaly discussed here and its obvious potential for future applications, we envision numerous possible extensions of this work. On the theory side there are multiple questions that one can ask: Which of the EM effects in solids translate to pseudo-EM fields discussed here? What are the best materials to study the effects? Do the chiral states predicted by our work  have prospects for designing more exotic many-body states in the presence of interactions? We also expect experimental activity to be stimulated since our predictions made for real materials yield effects that should be both unusual and eminently observable by conventional experimental probes such as charge  transport, ultrasonic attenuation and EM field emission.

\begin{acknowledgments}

The authors are indebted to P. Abbamonte, D.A. Bonn, D.M. Broun, W.N. Hardy, T. Liu, A. Rahmani  and N.P. Ong for illuminating discussions and  thank NSERC, CIfAR and Max Planck - UBC Centre for Quantum Materials for support.  Numerical simulations for Figs.\ 8 and 10 were performed using the Kwant code \cite{groth14} with computer resources provided by WestGrid and Compute Canada Calcul.
\end{acknowledgments}

\bibliography{weyl1.bib}

\begin{thebibliography}{47}%
\makeatletter
\providecommand \@ifxundefined [1]{%
 \@ifx{#1\undefined}
}%
\providecommand \@ifnum [1]{%
 \ifnum #1\expandafter \@firstoftwo
 \else \expandafter \@secondoftwo
 \fi
}%
\providecommand \@ifx [1]{%
 \ifx #1\expandafter \@firstoftwo
 \else \expandafter \@secondoftwo
 \fi
}%
\providecommand \natexlab [1]{#1}%
\providecommand \enquote  [1]{``#1''}%
\providecommand \bibnamefont  [1]{#1}%
\providecommand \bibfnamefont [1]{#1}%
\providecommand \citenamefont [1]{#1}%
\providecommand \href@noop [0]{\@secondoftwo}%
\providecommand \href [0]{\begingroup \@sanitize@url \@href}%
\providecommand \@href[1]{\@@startlink{#1}\@@href}%
\providecommand \@@href[1]{\endgroup#1\@@endlink}%
\providecommand \@sanitize@url [0]{\catcode `\\12\catcode `\$12\catcode
  `\&12\catcode `\#12\catcode `\^12\catcode `\_12\catcode `\%12\relax}%
\providecommand \@@startlink[1]{}%
\providecommand \@@endlink[0]{}%
\providecommand \url  [0]{\begingroup\@sanitize@url \@url }%
\providecommand \@url [1]{\endgroup\@href {#1}{\urlprefix }}%
\providecommand \urlprefix  [0]{URL }%
\providecommand \Eprint [0]{\href }%
\providecommand \doibase [0]{http://dx.doi.org/}%
\providecommand \selectlanguage [0]{\@gobble}%
\providecommand \bibinfo  [0]{\@secondoftwo}%
\providecommand \bibfield  [0]{\@secondoftwo}%
\providecommand \translation [1]{[#1]}%
\providecommand \BibitemOpen [0]{}%
\providecommand \bibitemStop [0]{}%
\providecommand \bibitemNoStop [0]{.\EOS\space}%
\providecommand \EOS [0]{\spacefactor3000\relax}%
\providecommand \BibitemShut  [1]{\csname bibitem#1\endcsname}%
\let\auto@bib@innerbib\@empty
%</preamble>
\bibitem [{\citenamefont {Guinea}\ \emph {et~al.}(2010)\citenamefont {Guinea},
  \citenamefont {Katsnelson},\ and\ \citenamefont {Geim}}]{guinea2010}%
  \BibitemOpen
  \bibfield  {author} {\bibinfo {author} {\bibfnamefont {F.}~\bibnamefont
  {Guinea}}, \bibinfo {author} {\bibfnamefont {M.~I.}\ \bibnamefont
  {Katsnelson}}, \ and\ \bibinfo {author} {\bibfnamefont {A.~K.}\ \bibnamefont
  {Geim}},\ }\bibfield  {title} {\enquote {\bibinfo {title} {Energy gaps and a
  zero-field quantum hall effect in graphene by strain engineering},}\ }\href
  {\doibase 10.1038/nphys1420} {\bibfield  {journal} {\bibinfo  {journal} {Nat
  Phys}\ }\textbf {\bibinfo {volume} {6}},\ \bibinfo {pages} {30--33} (\bibinfo
  {year} {2010})}\BibitemShut {NoStop}%
\bibitem [{\citenamefont {Levy}\ \emph {et~al.}(2010)\citenamefont {Levy},
  \citenamefont {Burke}, \citenamefont {Meaker}, \citenamefont {Panlasigui},
  \citenamefont {Zettl}, \citenamefont {Guinea}, \citenamefont {Neto},\ and\
  \citenamefont {Crommie}}]{levy2010}%
  \BibitemOpen
  \bibfield  {author} {\bibinfo {author} {\bibfnamefont {N.}~\bibnamefont
  {Levy}}, \bibinfo {author} {\bibfnamefont {S.~A.}\ \bibnamefont {Burke}},
  \bibinfo {author} {\bibfnamefont {K.~L.}\ \bibnamefont {Meaker}}, \bibinfo
  {author} {\bibfnamefont {M.}~\bibnamefont {Panlasigui}}, \bibinfo {author}
  {\bibfnamefont {A.}~\bibnamefont {Zettl}}, \bibinfo {author} {\bibfnamefont
  {F.}~\bibnamefont {Guinea}}, \bibinfo {author} {\bibfnamefont {A.~H.~Castro}\
  \bibnamefont {Neto}}, \ and\ \bibinfo {author} {\bibfnamefont {M.~F.}\
  \bibnamefont {Crommie}},\ }\bibfield  {title} {\enquote {\bibinfo {title}
  {Strain-induced pseudo{\textendash}magnetic fields greater than 300 tesla in
  graphene nanobubbles},}\ }\href {\doibase 10.1126/science.1191700} {\bibfield
   {journal} {\bibinfo  {journal} {Science}\ }\textbf {\bibinfo {volume}
  {329}},\ \bibinfo {pages} {544--547} (\bibinfo {year} {2010})}\BibitemShut
  {NoStop}%
\bibitem [{\citenamefont {Wan}\ \emph {et~al.}(2011)\citenamefont {Wan},
  \citenamefont {Turner}, \citenamefont {Vishwanath},\ and\ \citenamefont
  {Savrasov}}]{Savrasov2011}%
  \BibitemOpen
  \bibfield  {author} {\bibinfo {author} {\bibfnamefont {Xiangang}\
  \bibnamefont {Wan}}, \bibinfo {author} {\bibfnamefont {Ari~M.}\ \bibnamefont
  {Turner}}, \bibinfo {author} {\bibfnamefont {Ashvin}\ \bibnamefont
  {Vishwanath}}, \ and\ \bibinfo {author} {\bibfnamefont {Sergey~Y.}\
  \bibnamefont {Savrasov}},\ }\bibfield  {title} {\enquote {\bibinfo {title}
  {Topological semimetal and fermi-arc surface states in the electronic
  structure of pyrochlore iridates},}\ }\href {\doibase
  10.1103/PhysRevB.83.205101} {\bibfield  {journal} {\bibinfo  {journal} {Phys.
  Rev. B}\ }\textbf {\bibinfo {volume} {83}},\ \bibinfo {pages} {205101}
  (\bibinfo {year} {2011})}\BibitemShut {NoStop}%
\bibitem [{\citenamefont {Burkov}\ \emph {et~al.}(2011)\citenamefont {Burkov},
  \citenamefont {Hook},\ and\ \citenamefont {Balents}}]{burkov2011b}%
  \BibitemOpen
  \bibfield  {author} {\bibinfo {author} {\bibfnamefont {A.~A.}\ \bibnamefont
  {Burkov}}, \bibinfo {author} {\bibfnamefont {M.~D.}\ \bibnamefont {Hook}}, \
  and\ \bibinfo {author} {\bibfnamefont {Leon}\ \bibnamefont {Balents}},\
  }\bibfield  {title} {\enquote {\bibinfo {title} {Topological nodal
  semimetals},}\ }\href {\doibase 10.1103/PhysRevB.84.235126} {\bibfield
  {journal} {\bibinfo  {journal} {Phys. Rev. B}\ }\textbf {\bibinfo {volume}
  {84}},\ \bibinfo {pages} {235126} (\bibinfo {year} {2011})}\BibitemShut
  {NoStop}%
\bibitem [{\citenamefont {Vafek}\ and\ \citenamefont
  {Vishwanath}(2014)}]{Vafek2014}%
  \BibitemOpen
  \bibfield  {author} {\bibinfo {author} {\bibfnamefont {Oskar}\ \bibnamefont
  {Vafek}}\ and\ \bibinfo {author} {\bibfnamefont {Ashvin}\ \bibnamefont
  {Vishwanath}},\ }\bibfield  {title} {\enquote {\bibinfo {title} {Dirac
  fermions in solids: From high-tc cuprates and graphene to topological
  insulators and weyl semimetals},}\ }\href {\doibase
  10.1146/annurev-conmatphys-031113-133841} {\bibfield  {journal} {\bibinfo
  {journal} {Annual Review of Condensed Matter Physics}\ }\textbf {\bibinfo
  {volume} {5}},\ \bibinfo {pages} {83--112} (\bibinfo {year}
  {2014})}\BibitemShut {NoStop}%
\bibitem [{\citenamefont {Cortijo}\ \emph {et~al.}(2015)\citenamefont
  {Cortijo}, \citenamefont {Ferreir\'os}, \citenamefont {Landsteiner},\ and\
  \citenamefont {Vozmediano}}]{cortijo2015}%
  \BibitemOpen
  \bibfield  {author} {\bibinfo {author} {\bibfnamefont {Alberto}\ \bibnamefont
  {Cortijo}}, \bibinfo {author} {\bibfnamefont {Yago}\ \bibnamefont
  {Ferreir\'os}}, \bibinfo {author} {\bibfnamefont {Karl}\ \bibnamefont
  {Landsteiner}}, \ and\ \bibinfo {author} {\bibfnamefont {Mar\'{\i}a A.~H.}\
  \bibnamefont {Vozmediano}},\ }\bibfield  {title} {\enquote {\bibinfo {title}
  {Elastic gauge fields in weyl semimetals},}\ }\href {\doibase
  10.1103/PhysRevLett.115.177202} {\bibfield  {journal} {\bibinfo  {journal}
  {Phys. Rev. Lett.}\ }\textbf {\bibinfo {volume} {115}},\ \bibinfo {pages}
  {177202} (\bibinfo {year} {2015})}\BibitemShut {NoStop}%
\bibitem [{\citenamefont {Wang}\ \emph {et~al.}(2013)\citenamefont {Wang},
  \citenamefont {Weng}, \citenamefont {Wu}, \citenamefont {Dai},\ and\
  \citenamefont {Fang}}]{zhizhun2013}%
  \BibitemOpen
  \bibfield  {author} {\bibinfo {author} {\bibfnamefont {Zhijun}\ \bibnamefont
  {Wang}}, \bibinfo {author} {\bibfnamefont {Hongming}\ \bibnamefont {Weng}},
  \bibinfo {author} {\bibfnamefont {Quansheng}\ \bibnamefont {Wu}}, \bibinfo
  {author} {\bibfnamefont {Xi}~\bibnamefont {Dai}}, \ and\ \bibinfo {author}
  {\bibfnamefont {Zhong}\ \bibnamefont {Fang}},\ }\bibfield  {title} {\enquote
  {\bibinfo {title} {Three-dimensional dirac semimetal and quantum transport in
  cd${}_{3}$as${}_{2}$},}\ }\href {\doibase 10.1103/PhysRevB.88.125427}
  {\bibfield  {journal} {\bibinfo  {journal} {Phys. Rev. B}\ }\textbf {\bibinfo
  {volume} {88}},\ \bibinfo {pages} {125427} (\bibinfo {year}
  {2013})}\BibitemShut {NoStop}%
\bibitem [{\citenamefont {Borisenko}\ \emph {et~al.}(2014)\citenamefont
  {Borisenko}, \citenamefont {Gibson}, \citenamefont {Evtushinsky},
  \citenamefont {Zabolotnyy}, \citenamefont {B\"uchner},\ and\ \citenamefont
  {Cava}}]{borisenko2014}%
  \BibitemOpen
  \bibfield  {author} {\bibinfo {author} {\bibfnamefont {Sergey}\ \bibnamefont
  {Borisenko}}, \bibinfo {author} {\bibfnamefont {Quinn}\ \bibnamefont
  {Gibson}}, \bibinfo {author} {\bibfnamefont {Danil}\ \bibnamefont
  {Evtushinsky}}, \bibinfo {author} {\bibfnamefont {Volodymyr}\ \bibnamefont
  {Zabolotnyy}}, \bibinfo {author} {\bibfnamefont {Bernd}\ \bibnamefont
  {B\"uchner}}, \ and\ \bibinfo {author} {\bibfnamefont {Robert~J.}\
  \bibnamefont {Cava}},\ }\bibfield  {title} {\enquote {\bibinfo {title}
  {Experimental realization of a three-dimensional dirac semimetal},}\ }\href
  {\doibase 10.1103/PhysRevLett.113.027603} {\bibfield  {journal} {\bibinfo
  {journal} {Phys. Rev. Lett.}\ }\textbf {\bibinfo {volume} {113}},\ \bibinfo
  {pages} {027603} (\bibinfo {year} {2014})}\BibitemShut {NoStop}%
\bibitem [{\citenamefont {Neupane}\ \emph {et~al.}(2014)\citenamefont
  {Neupane}, \citenamefont {Xu}, \citenamefont {Sankar}, \citenamefont
  {Alidoust}, \citenamefont {Bian}, \citenamefont {Liu}, \citenamefont
  {Belopolski}, \citenamefont {Chang}, \citenamefont {Jeng}, \citenamefont
  {Lin}, \citenamefont {Bansil}, \citenamefont {Chou},\ and\ \citenamefont
  {Hasan}}]{neupane2014}%
  \BibitemOpen
  \bibfield  {author} {\bibinfo {author} {\bibfnamefont {Madhab}\ \bibnamefont
  {Neupane}}, \bibinfo {author} {\bibfnamefont {Su-Yang}\ \bibnamefont {Xu}},
  \bibinfo {author} {\bibfnamefont {Raman}\ \bibnamefont {Sankar}}, \bibinfo
  {author} {\bibfnamefont {Nasser}\ \bibnamefont {Alidoust}}, \bibinfo {author}
  {\bibfnamefont {Guang}\ \bibnamefont {Bian}}, \bibinfo {author}
  {\bibfnamefont {Chang}\ \bibnamefont {Liu}}, \bibinfo {author} {\bibfnamefont
  {Ilya}\ \bibnamefont {Belopolski}}, \bibinfo {author} {\bibfnamefont
  {Tay-Rong}\ \bibnamefont {Chang}}, \bibinfo {author} {\bibfnamefont
  {Horng-Tay}\ \bibnamefont {Jeng}}, \bibinfo {author} {\bibfnamefont {Hsin}\
  \bibnamefont {Lin}}, \bibinfo {author} {\bibfnamefont {Arun}\ \bibnamefont
  {Bansil}}, \bibinfo {author} {\bibfnamefont {Fangcheng}\ \bibnamefont
  {Chou}}, \ and\ \bibinfo {author} {\bibfnamefont {M.~Zahid}\ \bibnamefont
  {Hasan}},\ }\bibfield  {title} {\enquote {\bibinfo {title} {Observation of a
  three-dimensional topological dirac semimetal phase in high-mobility
  cd3as2},}\ }\href@noop {} {\bibfield  {journal} {\bibinfo  {journal} {Nat
  Commun}\ }\textbf {\bibinfo {volume} {5}},\ \bibinfo {pages} {4786} (\bibinfo
  {year} {2014})}\BibitemShut {NoStop}%
\bibitem [{\citenamefont {Jeon}\ \emph {et~al.}(2014)\citenamefont {Jeon},
  \citenamefont {Zhou}, \citenamefont {Gyenis}, \citenamefont {Feldman},
  \citenamefont {Kimchi}, \citenamefont {Potter}, \citenamefont {Gibson},
  \citenamefont {Cava}, \citenamefont {Vishwanath},\ and\ \citenamefont
  {Yazdani}}]{jeon2014}%
  \BibitemOpen
  \bibfield  {author} {\bibinfo {author} {\bibfnamefont {Sangjun}\ \bibnamefont
  {Jeon}}, \bibinfo {author} {\bibfnamefont {Brian~B.}\ \bibnamefont {Zhou}},
  \bibinfo {author} {\bibfnamefont {Andras}\ \bibnamefont {Gyenis}}, \bibinfo
  {author} {\bibfnamefont {Benjamin~E.}\ \bibnamefont {Feldman}}, \bibinfo
  {author} {\bibfnamefont {Itamar}\ \bibnamefont {Kimchi}}, \bibinfo {author}
  {\bibfnamefont {Andrew~C.}\ \bibnamefont {Potter}}, \bibinfo {author}
  {\bibfnamefont {Quinn~D.}\ \bibnamefont {Gibson}}, \bibinfo {author}
  {\bibfnamefont {Robert~J.}\ \bibnamefont {Cava}}, \bibinfo {author}
  {\bibfnamefont {Ashvin}\ \bibnamefont {Vishwanath}}, \ and\ \bibinfo {author}
  {\bibfnamefont {Ali}\ \bibnamefont {Yazdani}},\ }\bibfield  {title} {\enquote
  {\bibinfo {title} {Landau quantization and quasiparticle interference in the
  three-dimensional dirac?semimetal cd3as2},}\ }\href@noop {} {\bibfield
  {journal} {\bibinfo  {journal} {Nat Mater}\ }\textbf {\bibinfo {volume}
  {13}},\ \bibinfo {pages} {851--856} (\bibinfo {year} {2014})}\BibitemShut
  {NoStop}%
\bibitem [{\citenamefont {He}\ \emph {et~al.}(2014)\citenamefont {He},
  \citenamefont {Hong}, \citenamefont {Dong}, \citenamefont {Pan},
  \citenamefont {Zhang}, \citenamefont {Zhang},\ and\ \citenamefont
  {Li}}]{he2014}%
  \BibitemOpen
  \bibfield  {author} {\bibinfo {author} {\bibfnamefont {L.~P.}\ \bibnamefont
  {He}}, \bibinfo {author} {\bibfnamefont {X.~C.}\ \bibnamefont {Hong}},
  \bibinfo {author} {\bibfnamefont {J.~K.}\ \bibnamefont {Dong}}, \bibinfo
  {author} {\bibfnamefont {J.}~\bibnamefont {Pan}}, \bibinfo {author}
  {\bibfnamefont {Z.}~\bibnamefont {Zhang}}, \bibinfo {author} {\bibfnamefont
  {J.}~\bibnamefont {Zhang}}, \ and\ \bibinfo {author} {\bibfnamefont {S.~Y.}\
  \bibnamefont {Li}},\ }\bibfield  {title} {\enquote {\bibinfo {title} {Quantum
  transport evidence for the three-dimensional dirac semimetal phase in
  ${\mathrm{cd}}_{3}{\mathrm{as}}_{2}$},}\ }\href {\doibase
  10.1103/PhysRevLett.113.246402} {\bibfield  {journal} {\bibinfo  {journal}
  {Phys. Rev. Lett.}\ }\textbf {\bibinfo {volume} {113}},\ \bibinfo {pages}
  {246402} (\bibinfo {year} {2014})}\BibitemShut {NoStop}%
\bibitem [{\citenamefont {Liu}\ \emph {et~al.}(2014{\natexlab{a}})\citenamefont
  {Liu}, \citenamefont {Jiang}, \citenamefont {Zhou}, \citenamefont {Wang},
  \citenamefont {Zhang}, \citenamefont {Weng}, \citenamefont {Prabhakaran},
  \citenamefont {Mo}, \citenamefont {Peng}, \citenamefont {Dudin},
  \citenamefont {Kim}, \citenamefont {Hoesch}, \citenamefont {Fang},
  \citenamefont {Dai}, \citenamefont {Shen}, \citenamefont {Feng},
  \citenamefont {Hussain},\ and\ \citenamefont {Chen}}]{liu2014b}%
  \BibitemOpen
  \bibfield  {author} {\bibinfo {author} {\bibfnamefont {Z.~K.}\ \bibnamefont
  {Liu}}, \bibinfo {author} {\bibfnamefont {J.}~\bibnamefont {Jiang}}, \bibinfo
  {author} {\bibfnamefont {B.}~\bibnamefont {Zhou}}, \bibinfo {author}
  {\bibfnamefont {Z.~J.}\ \bibnamefont {Wang}}, \bibinfo {author}
  {\bibfnamefont {Y.}~\bibnamefont {Zhang}}, \bibinfo {author} {\bibfnamefont
  {H.~M.}\ \bibnamefont {Weng}}, \bibinfo {author} {\bibfnamefont
  {D.}~\bibnamefont {Prabhakaran}}, \bibinfo {author} {\bibfnamefont {S.-K.}\
  \bibnamefont {Mo}}, \bibinfo {author} {\bibfnamefont {H.}~\bibnamefont
  {Peng}}, \bibinfo {author} {\bibfnamefont {P.}~\bibnamefont {Dudin}},
  \bibinfo {author} {\bibfnamefont {T.}~\bibnamefont {Kim}}, \bibinfo {author}
  {\bibfnamefont {M.}~\bibnamefont {Hoesch}}, \bibinfo {author} {\bibfnamefont
  {Z.}~\bibnamefont {Fang}}, \bibinfo {author} {\bibfnamefont {X.}~\bibnamefont
  {Dai}}, \bibinfo {author} {\bibfnamefont {Z.~X.}\ \bibnamefont {Shen}},
  \bibinfo {author} {\bibfnamefont {D.~L.}\ \bibnamefont {Feng}}, \bibinfo
  {author} {\bibfnamefont {Z.}~\bibnamefont {Hussain}}, \ and\ \bibinfo
  {author} {\bibfnamefont {Y.~L.}\ \bibnamefont {Chen}},\ }\bibfield  {title}
  {\enquote {\bibinfo {title} {A stable three-dimensional topological dirac
  semimetal cd3as2},}\ }\href@noop {} {\bibfield  {journal} {\bibinfo
  {journal} {Nat Mater}\ }\textbf {\bibinfo {volume} {13}},\ \bibinfo {pages}
  {677--681} (\bibinfo {year} {2014}{\natexlab{a}})}\BibitemShut {NoStop}%
\bibitem [{\citenamefont {Wang}\ \emph {et~al.}(2012)\citenamefont {Wang},
  \citenamefont {Sun}, \citenamefont {Chen}, \citenamefont {Franchini},
  \citenamefont {Xu}, \citenamefont {Weng}, \citenamefont {Dai},\ and\
  \citenamefont {Fang}}]{zhizhun2012}%
  \BibitemOpen
  \bibfield  {author} {\bibinfo {author} {\bibfnamefont {Zhijun}\ \bibnamefont
  {Wang}}, \bibinfo {author} {\bibfnamefont {Yan}\ \bibnamefont {Sun}},
  \bibinfo {author} {\bibfnamefont {Xing-Qiu}\ \bibnamefont {Chen}}, \bibinfo
  {author} {\bibfnamefont {Cesare}\ \bibnamefont {Franchini}}, \bibinfo
  {author} {\bibfnamefont {Gang}\ \bibnamefont {Xu}}, \bibinfo {author}
  {\bibfnamefont {Hongming}\ \bibnamefont {Weng}}, \bibinfo {author}
  {\bibfnamefont {Xi}~\bibnamefont {Dai}}, \ and\ \bibinfo {author}
  {\bibfnamefont {Zhong}\ \bibnamefont {Fang}},\ }\bibfield  {title} {\enquote
  {\bibinfo {title} {Dirac semimetal and topological phase transitions in
  ${A}_{3}$bi ($a=\text{Na}$, k, rb)},}\ }\href {\doibase
  10.1103/PhysRevB.85.195320} {\bibfield  {journal} {\bibinfo  {journal} {Phys.
  Rev. B}\ }\textbf {\bibinfo {volume} {85}},\ \bibinfo {pages} {195320}
  (\bibinfo {year} {2012})}\BibitemShut {NoStop}%
\bibitem [{\citenamefont {Liu}\ \emph {et~al.}(2014{\natexlab{b}})\citenamefont
  {Liu}, \citenamefont {Zhou}, \citenamefont {Zhang}, \citenamefont {Wang},
  \citenamefont {Weng}, \citenamefont {Prabhakaran}, \citenamefont {Mo},
  \citenamefont {Shen}, \citenamefont {Fang}, \citenamefont {Dai},
  \citenamefont {Hussain},\ and\ \citenamefont {Chen}}]{yulin2014}%
  \BibitemOpen
  \bibfield  {author} {\bibinfo {author} {\bibfnamefont {Z.~K.}\ \bibnamefont
  {Liu}}, \bibinfo {author} {\bibfnamefont {B.}~\bibnamefont {Zhou}}, \bibinfo
  {author} {\bibfnamefont {Y.}~\bibnamefont {Zhang}}, \bibinfo {author}
  {\bibfnamefont {Z.~J.}\ \bibnamefont {Wang}}, \bibinfo {author}
  {\bibfnamefont {H.~M.}\ \bibnamefont {Weng}}, \bibinfo {author}
  {\bibfnamefont {D.}~\bibnamefont {Prabhakaran}}, \bibinfo {author}
  {\bibfnamefont {S.-K.}\ \bibnamefont {Mo}}, \bibinfo {author} {\bibfnamefont
  {Z.~X.}\ \bibnamefont {Shen}}, \bibinfo {author} {\bibfnamefont
  {Z.}~\bibnamefont {Fang}}, \bibinfo {author} {\bibfnamefont {X.}~\bibnamefont
  {Dai}}, \bibinfo {author} {\bibfnamefont {Z.}~\bibnamefont {Hussain}}, \ and\
  \bibinfo {author} {\bibfnamefont {Y.~L.}\ \bibnamefont {Chen}},\ }\bibfield
  {title} {\enquote {\bibinfo {title} {Discovery of a three-dimensional
  topological dirac semimetal, na3bi},}\ }\href {\doibase
  10.1126/science.1245085} {\bibfield  {journal} {\bibinfo  {journal}
  {Science}\ }\textbf {\bibinfo {volume} {343}},\ \bibinfo {pages} {864--867}
  (\bibinfo {year} {2014}{\natexlab{b}})}\BibitemShut {NoStop}%
\bibitem [{\citenamefont {Zhang}\ \emph {et~al.}(2014)\citenamefont {Zhang},
  \citenamefont {Liu}, \citenamefont {Zhou}, \citenamefont {Kim}, \citenamefont
  {Hussain}, \citenamefont {Shen}, \citenamefont {Chen},\ and\ \citenamefont
  {Mo}}]{yulin2014b}%
  \BibitemOpen
  \bibfield  {author} {\bibinfo {author} {\bibfnamefont {Yi}~\bibnamefont
  {Zhang}}, \bibinfo {author} {\bibfnamefont {Zhongkai}\ \bibnamefont {Liu}},
  \bibinfo {author} {\bibfnamefont {Bo}~\bibnamefont {Zhou}}, \bibinfo {author}
  {\bibfnamefont {Yeongkwan}\ \bibnamefont {Kim}}, \bibinfo {author}
  {\bibfnamefont {Zahid}\ \bibnamefont {Hussain}}, \bibinfo {author}
  {\bibfnamefont {Zhi-Xun}\ \bibnamefont {Shen}}, \bibinfo {author}
  {\bibfnamefont {Yulin}\ \bibnamefont {Chen}}, \ and\ \bibinfo {author}
  {\bibfnamefont {Sung-Kwan}\ \bibnamefont {Mo}},\ }\bibfield  {title}
  {\enquote {\bibinfo {title} {Molecular beam epitaxial growth of a
  three-dimensional topological dirac semimetal na3bi},}\ }\href@noop {}
  {\bibfield  {journal} {\bibinfo  {journal} {Applied Physics Letters}\
  }\textbf {\bibinfo {volume} {105}},\ \bibinfo {eid} {031901} (\bibinfo {year}
  {2014})}\BibitemShut {NoStop}%
\bibitem [{\citenamefont {Xu}\ \emph {et~al.}(2015{\natexlab{a}})\citenamefont
  {Xu}, \citenamefont {Belopolski}, \citenamefont {Alidoust}, \citenamefont
  {Neupane}, \citenamefont {Bian}, \citenamefont {Zhang}, \citenamefont
  {Sankar}, \citenamefont {Chang}, \citenamefont {Yuan}, \citenamefont {Lee},
  \citenamefont {Huang}, \citenamefont {Zheng}, \citenamefont {Ma},
  \citenamefont {Sanchez}, \citenamefont {Wang}, \citenamefont {Bansil},
  \citenamefont {Chou}, \citenamefont {Shibayev}, \citenamefont {Lin},
  \citenamefont {Jia},\ and\ \citenamefont {Hasan}}]{hasan2015}%
  \BibitemOpen
  \bibfield  {author} {\bibinfo {author} {\bibfnamefont {Su-Yang}\ \bibnamefont
  {Xu}}, \bibinfo {author} {\bibfnamefont {Ilya}\ \bibnamefont {Belopolski}},
  \bibinfo {author} {\bibfnamefont {Nasser}\ \bibnamefont {Alidoust}}, \bibinfo
  {author} {\bibfnamefont {Madhab}\ \bibnamefont {Neupane}}, \bibinfo {author}
  {\bibfnamefont {Guang}\ \bibnamefont {Bian}}, \bibinfo {author}
  {\bibfnamefont {Chenglong}\ \bibnamefont {Zhang}}, \bibinfo {author}
  {\bibfnamefont {Raman}\ \bibnamefont {Sankar}}, \bibinfo {author}
  {\bibfnamefont {Guoqing}\ \bibnamefont {Chang}}, \bibinfo {author}
  {\bibfnamefont {Zhujun}\ \bibnamefont {Yuan}}, \bibinfo {author}
  {\bibfnamefont {Chi-Cheng}\ \bibnamefont {Lee}}, \bibinfo {author}
  {\bibfnamefont {Shin-Ming}\ \bibnamefont {Huang}}, \bibinfo {author}
  {\bibfnamefont {Hao}\ \bibnamefont {Zheng}}, \bibinfo {author} {\bibfnamefont
  {Jie}\ \bibnamefont {Ma}}, \bibinfo {author} {\bibfnamefont {Daniel~S.}\
  \bibnamefont {Sanchez}}, \bibinfo {author} {\bibfnamefont {BaoKai}\
  \bibnamefont {Wang}}, \bibinfo {author} {\bibfnamefont {Arun}\ \bibnamefont
  {Bansil}}, \bibinfo {author} {\bibfnamefont {Fangcheng}\ \bibnamefont
  {Chou}}, \bibinfo {author} {\bibfnamefont {Pavel~P.}\ \bibnamefont
  {Shibayev}}, \bibinfo {author} {\bibfnamefont {Hsin}\ \bibnamefont {Lin}},
  \bibinfo {author} {\bibfnamefont {Shuang}\ \bibnamefont {Jia}}, \ and\
  \bibinfo {author} {\bibfnamefont {M.~Zahid}\ \bibnamefont {Hasan}},\
  }\bibfield  {title} {\enquote {\bibinfo {title} {Discovery of a weyl fermion
  semimetal and topological fermi arcs},}\ }\href {\doibase
  10.1126/science.aaa9297} {\bibfield  {journal} {\bibinfo  {journal}
  {Science}\ }\textbf {\bibinfo {volume} {349}},\ \bibinfo {pages} {613--617}
  (\bibinfo {year} {2015}{\natexlab{a}})}\BibitemShut {NoStop}%
\bibitem [{\citenamefont {Lv}\ \emph {et~al.}(2015)\citenamefont {Lv},
  \citenamefont {Weng}, \citenamefont {Fu}, \citenamefont {Wang}, \citenamefont
  {Miao}, \citenamefont {Ma}, \citenamefont {Richard}, \citenamefont {Huang},
  \citenamefont {Zhao}, \citenamefont {Chen}, \citenamefont {Fang},
  \citenamefont {Dai}, \citenamefont {Qian},\ and\ \citenamefont
  {Ding}}]{ding2015}%
  \BibitemOpen
  \bibfield  {author} {\bibinfo {author} {\bibfnamefont {B.~Q.}\ \bibnamefont
  {Lv}}, \bibinfo {author} {\bibfnamefont {H.~M.}\ \bibnamefont {Weng}},
  \bibinfo {author} {\bibfnamefont {B.~B.}\ \bibnamefont {Fu}}, \bibinfo
  {author} {\bibfnamefont {X.~P.}\ \bibnamefont {Wang}}, \bibinfo {author}
  {\bibfnamefont {H.}~\bibnamefont {Miao}}, \bibinfo {author} {\bibfnamefont
  {J.}~\bibnamefont {Ma}}, \bibinfo {author} {\bibfnamefont {P.}~\bibnamefont
  {Richard}}, \bibinfo {author} {\bibfnamefont {X.~C.}\ \bibnamefont {Huang}},
  \bibinfo {author} {\bibfnamefont {L.~X.}\ \bibnamefont {Zhao}}, \bibinfo
  {author} {\bibfnamefont {G.~F.}\ \bibnamefont {Chen}}, \bibinfo {author}
  {\bibfnamefont {Z.}~\bibnamefont {Fang}}, \bibinfo {author} {\bibfnamefont
  {X.}~\bibnamefont {Dai}}, \bibinfo {author} {\bibfnamefont {T.}~\bibnamefont
  {Qian}}, \ and\ \bibinfo {author} {\bibfnamefont {H.}~\bibnamefont {Ding}},\
  }\bibfield  {title} {\enquote {\bibinfo {title} {Experimental discovery of
  weyl semimetal taas},}\ }\href {\doibase 10.1103/PhysRevX.5.031013}
  {\bibfield  {journal} {\bibinfo  {journal} {Phys. Rev. X}\ }\textbf {\bibinfo
  {volume} {5}},\ \bibinfo {pages} {031013} (\bibinfo {year}
  {2015})}\BibitemShut {NoStop}%
\bibitem [{\citenamefont {Shekhar}\ \emph {et~al.}(2015)\citenamefont
  {Shekhar}, \citenamefont {Nayak}, \citenamefont {Sun}, \citenamefont
  {Schmidt}, \citenamefont {Nicklas}, \citenamefont {Leermakers}, \citenamefont
  {Zeitler}, \citenamefont {Skourski}, \citenamefont {Wosnitza}, \citenamefont
  {Liu}, \citenamefont {Chen}, \citenamefont {Schnelle}, \citenamefont
  {Borrmann}, \citenamefont {Grin}, \citenamefont {Felser},\ and\ \citenamefont
  {Yan}}]{yan2015}%
  \BibitemOpen
  \bibfield  {author} {\bibinfo {author} {\bibfnamefont {Chandra}\ \bibnamefont
  {Shekhar}}, \bibinfo {author} {\bibfnamefont {Ajaya~K.}\ \bibnamefont
  {Nayak}}, \bibinfo {author} {\bibfnamefont {Yan}\ \bibnamefont {Sun}},
  \bibinfo {author} {\bibfnamefont {Marcus}\ \bibnamefont {Schmidt}}, \bibinfo
  {author} {\bibfnamefont {Michael}\ \bibnamefont {Nicklas}}, \bibinfo {author}
  {\bibfnamefont {Inge}\ \bibnamefont {Leermakers}}, \bibinfo {author}
  {\bibfnamefont {Uli}\ \bibnamefont {Zeitler}}, \bibinfo {author}
  {\bibfnamefont {Yurii}\ \bibnamefont {Skourski}}, \bibinfo {author}
  {\bibfnamefont {Jochen}\ \bibnamefont {Wosnitza}}, \bibinfo {author}
  {\bibfnamefont {Zhongkai}\ \bibnamefont {Liu}}, \bibinfo {author}
  {\bibfnamefont {Yulin}\ \bibnamefont {Chen}}, \bibinfo {author}
  {\bibfnamefont {Walter}\ \bibnamefont {Schnelle}}, \bibinfo {author}
  {\bibfnamefont {Horst}\ \bibnamefont {Borrmann}}, \bibinfo {author}
  {\bibfnamefont {Yuri}\ \bibnamefont {Grin}}, \bibinfo {author} {\bibfnamefont
  {Claudia}\ \bibnamefont {Felser}}, \ and\ \bibinfo {author} {\bibfnamefont
  {Binghai}\ \bibnamefont {Yan}},\ }\bibfield  {title} {\enquote {\bibinfo
  {title} {Extremely large magnetoresistance and ultrahigh mobility in the
  topological weyl semimetal candidate nbp},}\ }\href@noop {} {\bibfield
  {journal} {\bibinfo  {journal} {Nat Phys}\ }\textbf {\bibinfo {volume}
  {11}},\ \bibinfo {pages} {645--649} (\bibinfo {year} {2015})},\ \bibinfo
  {note} {letter}\BibitemShut {NoStop}%
\bibitem [{\citenamefont {Yang}\ \emph {et~al.}(2015)\citenamefont {Yang},
  \citenamefont {Liu}, \citenamefont {Sun}, \citenamefont {Peng}, \citenamefont
  {Yang}, \citenamefont {Zhang}, \citenamefont {Zhou}, \citenamefont {Zhang},
  \citenamefont {Guo}, \citenamefont {Rahn}, \citenamefont {Prabhakaran},
  \citenamefont {Hussain}, \citenamefont {Mo}, \citenamefont {Felser},
  \citenamefont {Yan},\ and\ \citenamefont {Chen}}]{chen2015}%
  \BibitemOpen
  \bibfield  {author} {\bibinfo {author} {\bibfnamefont {L.~X.}\ \bibnamefont
  {Yang}}, \bibinfo {author} {\bibfnamefont {Z.~K.}\ \bibnamefont {Liu}},
  \bibinfo {author} {\bibfnamefont {Y.}~\bibnamefont {Sun}}, \bibinfo {author}
  {\bibfnamefont {H.}~\bibnamefont {Peng}}, \bibinfo {author} {\bibfnamefont
  {H.~F.}\ \bibnamefont {Yang}}, \bibinfo {author} {\bibfnamefont
  {T.}~\bibnamefont {Zhang}}, \bibinfo {author} {\bibfnamefont
  {B.}~\bibnamefont {Zhou}}, \bibinfo {author} {\bibfnamefont {Y.}~\bibnamefont
  {Zhang}}, \bibinfo {author} {\bibfnamefont {Y.~F.}\ \bibnamefont {Guo}},
  \bibinfo {author} {\bibfnamefont {M.}~\bibnamefont {Rahn}}, \bibinfo {author}
  {\bibfnamefont {D.}~\bibnamefont {Prabhakaran}}, \bibinfo {author}
  {\bibfnamefont {Z.}~\bibnamefont {Hussain}}, \bibinfo {author} {\bibfnamefont
  {S.-K.}\ \bibnamefont {Mo}}, \bibinfo {author} {\bibfnamefont
  {C.}~\bibnamefont {Felser}}, \bibinfo {author} {\bibfnamefont
  {B.}~\bibnamefont {Yan}}, \ and\ \bibinfo {author} {\bibfnamefont {Y.~L.}\
  \bibnamefont {Chen}},\ }\bibfield  {title} {\enquote {\bibinfo {title} {Weyl
  semimetal phase in the non-centrosymmetric compound taas},}\ }\href@noop {}
  {\bibfield  {journal} {\bibinfo  {journal} {Nat Phys}\ }\textbf {\bibinfo
  {volume} {11}},\ \bibinfo {pages} {728--732} (\bibinfo {year} {2015})},\
  \bibinfo {note} {letter}\BibitemShut {NoStop}%
\bibitem [{\citenamefont {Xu}\ \emph {et~al.}(2015{\natexlab{b}})\citenamefont
  {Xu}, \citenamefont {Alidoust}, \citenamefont {Belopolski}, \citenamefont
  {Yuan}, \citenamefont {Bian}, \citenamefont {Chang}, \citenamefont {Zheng},
  \citenamefont {Strocov}, \citenamefont {Sanchez}, \citenamefont {Chang},
  \citenamefont {Zhang}, \citenamefont {Mou}, \citenamefont {Wu}, \citenamefont
  {Huang}, \citenamefont {Lee}, \citenamefont {Huang}, \citenamefont {Wang},
  \citenamefont {Bansil}, \citenamefont {Jeng}, \citenamefont {Neupert},
  \citenamefont {Kaminski}, \citenamefont {Lin}, \citenamefont {Jia},\ and\
  \citenamefont {Zahid~Hasan}}]{xu2015}%
  \BibitemOpen
  \bibfield  {author} {\bibinfo {author} {\bibfnamefont {Su-Yang}\ \bibnamefont
  {Xu}}, \bibinfo {author} {\bibfnamefont {Nasser}\ \bibnamefont {Alidoust}},
  \bibinfo {author} {\bibfnamefont {Ilya}\ \bibnamefont {Belopolski}}, \bibinfo
  {author} {\bibfnamefont {Zhujun}\ \bibnamefont {Yuan}}, \bibinfo {author}
  {\bibfnamefont {Guang}\ \bibnamefont {Bian}}, \bibinfo {author}
  {\bibfnamefont {Tay-Rong}\ \bibnamefont {Chang}}, \bibinfo {author}
  {\bibfnamefont {Hao}\ \bibnamefont {Zheng}}, \bibinfo {author} {\bibfnamefont
  {Vladimir~N.}\ \bibnamefont {Strocov}}, \bibinfo {author} {\bibfnamefont
  {Daniel~S.}\ \bibnamefont {Sanchez}}, \bibinfo {author} {\bibfnamefont
  {Guoqing}\ \bibnamefont {Chang}}, \bibinfo {author} {\bibfnamefont
  {Chenglong}\ \bibnamefont {Zhang}}, \bibinfo {author} {\bibfnamefont
  {Daixiang}\ \bibnamefont {Mou}}, \bibinfo {author} {\bibfnamefont {Yun}\
  \bibnamefont {Wu}}, \bibinfo {author} {\bibfnamefont {Lunan}\ \bibnamefont
  {Huang}}, \bibinfo {author} {\bibfnamefont {Chi-Cheng}\ \bibnamefont {Lee}},
  \bibinfo {author} {\bibfnamefont {Shin-Ming}\ \bibnamefont {Huang}}, \bibinfo
  {author} {\bibfnamefont {BaoKai}\ \bibnamefont {Wang}}, \bibinfo {author}
  {\bibfnamefont {Arun}\ \bibnamefont {Bansil}}, \bibinfo {author}
  {\bibfnamefont {Horng-Tay}\ \bibnamefont {Jeng}}, \bibinfo {author}
  {\bibfnamefont {Titus}\ \bibnamefont {Neupert}}, \bibinfo {author}
  {\bibfnamefont {Adam}\ \bibnamefont {Kaminski}}, \bibinfo {author}
  {\bibfnamefont {Hsin}\ \bibnamefont {Lin}}, \bibinfo {author} {\bibfnamefont
  {Shuang}\ \bibnamefont {Jia}}, \ and\ \bibinfo {author} {\bibfnamefont
  {M.}~\bibnamefont {Zahid~Hasan}},\ }\bibfield  {title} {\enquote {\bibinfo
  {title} {Discovery of a weyl fermion state with fermi arcs in niobium
  arsenide},}\ }\href@noop {} {\bibfield  {journal} {\bibinfo  {journal} {Nat
  Phys}\ }\textbf {\bibinfo {volume} {11}},\ \bibinfo {pages} {748--754}
  (\bibinfo {year} {2015}{\natexlab{b}})},\ \bibinfo {note}
  {article}\BibitemShut {NoStop}%
\bibitem [{\citenamefont {Adler}(1969)}]{adler1969}%
  \BibitemOpen
  \bibfield  {author} {\bibinfo {author} {\bibfnamefont {Stephen~L.}\
  \bibnamefont {Adler}},\ }\bibfield  {title} {\enquote {\bibinfo {title}
  {Axial-vector vertex in spinor electrodynamics},}\ }\href {\doibase
  10.1103/PhysRev.177.2426} {\bibfield  {journal} {\bibinfo  {journal} {Phys.
  Rev.}\ }\textbf {\bibinfo {volume} {177}},\ \bibinfo {pages} {2426--2438}
  (\bibinfo {year} {1969})}\BibitemShut {NoStop}%
\bibitem [{\citenamefont {Bell}\ and\ \citenamefont {Jackiw}(1969)}]{bell1969}%
  \BibitemOpen
  \bibfield  {author} {\bibinfo {author} {\bibfnamefont {J.~S.}\ \bibnamefont
  {Bell}}\ and\ \bibinfo {author} {\bibfnamefont {R.}~\bibnamefont {Jackiw}},\
  }\bibfield  {title} {\enquote {\bibinfo {title} {A pcac puzzle:
  $\pi\to\gamma\gamma$ in the $\sigma$-model},}\ }\href {\doibase
  10.1007/BF02823296} {\bibfield  {journal} {\bibinfo  {journal} {Il Nuovo
  Cimento A (1971-1996)}\ }\textbf {\bibinfo {volume} {60}},\ \bibinfo {pages}
  {47--61} (\bibinfo {year} {1969})}\BibitemShut {NoStop}%
\bibitem [{\citenamefont {Nielsen}\ and\ \citenamefont
  {Ninomiya}(1983)}]{nielsen1983}%
  \BibitemOpen
  \bibfield  {author} {\bibinfo {author} {\bibfnamefont {H.B.}\ \bibnamefont
  {Nielsen}}\ and\ \bibinfo {author} {\bibfnamefont {Masao}\ \bibnamefont
  {Ninomiya}},\ }\bibfield  {title} {\enquote {\bibinfo {title} {The
  adler-bell-jackiw anomaly and weyl fermions in a crystal},}\ }\href@noop {}
  {\bibfield  {journal} {\bibinfo  {journal} {Physics Letters B}\ }\textbf
  {\bibinfo {volume} {130}},\ \bibinfo {pages} {389 -- 396} (\bibinfo {year}
  {1983})}\BibitemShut {NoStop}%
\bibitem [{\citenamefont {Li}\ \emph {et~al.}(2015)\citenamefont {Li},
  \citenamefont {Wang}, \citenamefont {Liu}, \citenamefont {Wang},
  \citenamefont {Liao},\ and\ \citenamefont {Yu}}]{li2015}%
  \BibitemOpen
  \bibfield  {author} {\bibinfo {author} {\bibfnamefont {Cai-Zhen}\
  \bibnamefont {Li}}, \bibinfo {author} {\bibfnamefont {Li-Xian}\ \bibnamefont
  {Wang}}, \bibinfo {author} {\bibfnamefont {Haiwen}\ \bibnamefont {Liu}},
  \bibinfo {author} {\bibfnamefont {Jian}\ \bibnamefont {Wang}}, \bibinfo
  {author} {\bibfnamefont {Zhi-Min}\ \bibnamefont {Liao}}, \ and\ \bibinfo
  {author} {\bibfnamefont {Da-Peng}\ \bibnamefont {Yu}},\ }\bibfield  {title}
  {\enquote {\bibinfo {title} {Giant negative magnetoresistance induced by the
  chiral anomaly in individual cd3as2 nanowires},}\ }\href@noop {} {\bibfield
  {journal} {\bibinfo  {journal} {Nat Commun}\ }\textbf {\bibinfo {volume}
  {6}},\ \bibinfo {pages} {10137} (\bibinfo {year} {2015})}\BibitemShut
  {NoStop}%
\bibitem [{\citenamefont {Wang}\ \emph {et~al.}(2016)\citenamefont {Wang},
  \citenamefont {Li}, \citenamefont {Yu},\ and\ \citenamefont
  {Liao}}]{wang2016}%
  \BibitemOpen
  \bibfield  {author} {\bibinfo {author} {\bibfnamefont {Li-Xian}\ \bibnamefont
  {Wang}}, \bibinfo {author} {\bibfnamefont {Cai-Zhen}\ \bibnamefont {Li}},
  \bibinfo {author} {\bibfnamefont {Da-Peng}\ \bibnamefont {Yu}}, \ and\
  \bibinfo {author} {\bibfnamefont {Zhi-Min}\ \bibnamefont {Liao}},\ }\bibfield
   {title} {\enquote {\bibinfo {title} {Aharonov-bohm oscillations in dirac
  semimetal cd3as2 nanowires},}\ }\href@noop {} {\bibfield  {journal} {\bibinfo
   {journal} {Nat Commun}\ }\textbf {\bibinfo {volume} {7}},\ \bibinfo {pages}
  {10769} (\bibinfo {year} {2016})}\BibitemShut {NoStop}%
\bibitem [{\citenamefont {Burkov}\ and\ \citenamefont
  {Balents}(2011)}]{burkov2011}%
  \BibitemOpen
  \bibfield  {author} {\bibinfo {author} {\bibfnamefont {A.~A.}\ \bibnamefont
  {Burkov}}\ and\ \bibinfo {author} {\bibfnamefont {Leon}\ \bibnamefont
  {Balents}},\ }\bibfield  {title} {\enquote {\bibinfo {title} {Weyl semimetal
  in a topological insulator multilayer},}\ }\href {\doibase
  10.1103/PhysRevLett.107.127205} {\bibfield  {journal} {\bibinfo  {journal}
  {Phys. Rev. Lett.}\ }\textbf {\bibinfo {volume} {107}},\ \bibinfo {pages}
  {127205} (\bibinfo {year} {2011})}\BibitemShut {NoStop}%
\bibitem [{\citenamefont {Shapourian}\ \emph {et~al.}(2015)\citenamefont
  {Shapourian}, \citenamefont {Hughes},\ and\ \citenamefont
  {Ryu}}]{shapourian2015}%
  \BibitemOpen
  \bibfield  {author} {\bibinfo {author} {\bibfnamefont {Hassan}\ \bibnamefont
  {Shapourian}}, \bibinfo {author} {\bibfnamefont {Taylor~L.}\ \bibnamefont
  {Hughes}}, \ and\ \bibinfo {author} {\bibfnamefont {Shinsei}\ \bibnamefont
  {Ryu}},\ }\bibfield  {title} {\enquote {\bibinfo {title} {Viscoelastic
  response of topological tight-binding models in two and three dimensions},}\
  }\href {\doibase 10.1103/PhysRevB.92.165131} {\bibfield  {journal} {\bibinfo
  {journal} {Phys. Rev. B}\ }\textbf {\bibinfo {volume} {92}},\ \bibinfo
  {pages} {165131} (\bibinfo {year} {2015})}\BibitemShut {NoStop}%
\bibitem [{\citenamefont {Liu}\ \emph {et~al.}(2013)\citenamefont {Liu},
  \citenamefont {Ye},\ and\ \citenamefont {Qi}}]{qi2013}%
  \BibitemOpen
  \bibfield  {author} {\bibinfo {author} {\bibfnamefont {Chao-Xing}\
  \bibnamefont {Liu}}, \bibinfo {author} {\bibfnamefont {Peng}\ \bibnamefont
  {Ye}}, \ and\ \bibinfo {author} {\bibfnamefont {Xiao-Liang}\ \bibnamefont
  {Qi}},\ }\bibfield  {title} {\enquote {\bibinfo {title} {Chiral gauge field
  and axial anomaly in a weyl semimetal},}\ }\href {\doibase
  10.1103/PhysRevB.87.235306} {\bibfield  {journal} {\bibinfo  {journal} {Phys.
  Rev. B}\ }\textbf {\bibinfo {volume} {87}},\ \bibinfo {pages} {235306}
  (\bibinfo {year} {2013})}\BibitemShut {NoStop}%
\bibitem [{\citenamefont {Fukushima}\ \emph {et~al.}(2008)\citenamefont
  {Fukushima}, \citenamefont {Kharzeev},\ and\ \citenamefont
  {Warringa}}]{fukushima2008}%
  \BibitemOpen
  \bibfield  {author} {\bibinfo {author} {\bibfnamefont {Kenji}\ \bibnamefont
  {Fukushima}}, \bibinfo {author} {\bibfnamefont {Dmitri~E.}\ \bibnamefont
  {Kharzeev}}, \ and\ \bibinfo {author} {\bibfnamefont {Harmen~J.}\
  \bibnamefont {Warringa}},\ }\bibfield  {title} {\enquote {\bibinfo {title}
  {Chiral magnetic effect},}\ }\href {\doibase 10.1103/PhysRevD.78.074033}
  {\bibfield  {journal} {\bibinfo  {journal} {Phys. Rev. D}\ }\textbf {\bibinfo
  {volume} {78}},\ \bibinfo {pages} {074033} (\bibinfo {year}
  {2008})}\BibitemShut {NoStop}%
\bibitem [{\citenamefont {Son}\ and\ \citenamefont {Spivak}(2013)}]{son2013}%
  \BibitemOpen
  \bibfield  {author} {\bibinfo {author} {\bibfnamefont {D.~T.}\ \bibnamefont
  {Son}}\ and\ \bibinfo {author} {\bibfnamefont {B.~Z.}\ \bibnamefont
  {Spivak}},\ }\bibfield  {title} {\enquote {\bibinfo {title} {Chiral anomaly
  and classical negative magnetoresistance of weyl metals},}\ }\href {\doibase
  10.1103/PhysRevB.88.104412} {\bibfield  {journal} {\bibinfo  {journal} {Phys.
  Rev. B}\ }\textbf {\bibinfo {volume} {88}},\ \bibinfo {pages} {104412}
  (\bibinfo {year} {2013})}\BibitemShut {NoStop}%
\bibitem [{\citenamefont {Burkov}(2015)}]{burkov2015}%
  \BibitemOpen
  \bibfield  {author} {\bibinfo {author} {\bibfnamefont {A~A}\ \bibnamefont
  {Burkov}},\ }\bibfield  {title} {\enquote {\bibinfo {title} {Chiral anomaly
  and transport in weyl metals},}\ }\href@noop {} {\bibfield  {journal}
  {\bibinfo  {journal} {Journal of Physics: Condensed Matter}\ }\textbf
  {\bibinfo {volume} {27}},\ \bibinfo {pages} {113201} (\bibinfo {year}
  {2015})}\BibitemShut {NoStop}%
\bibitem [{\citenamefont {Kim}\ \emph {et~al.}(2013)\citenamefont {Kim},
  \citenamefont {Kim}, \citenamefont {Wang}, \citenamefont {Sasaki},
  \citenamefont {Satoh}, \citenamefont {Ohnishi}, \citenamefont {Kitaura},
  \citenamefont {Yang},\ and\ \citenamefont {Li}}]{kim2013}%
  \BibitemOpen
  \bibfield  {author} {\bibinfo {author} {\bibfnamefont {Heon-Jung}\
  \bibnamefont {Kim}}, \bibinfo {author} {\bibfnamefont {Ki-Seok}\ \bibnamefont
  {Kim}}, \bibinfo {author} {\bibfnamefont {J.-F.}\ \bibnamefont {Wang}},
  \bibinfo {author} {\bibfnamefont {M.}~\bibnamefont {Sasaki}}, \bibinfo
  {author} {\bibfnamefont {N.}~\bibnamefont {Satoh}}, \bibinfo {author}
  {\bibfnamefont {A.}~\bibnamefont {Ohnishi}}, \bibinfo {author} {\bibfnamefont
  {M.}~\bibnamefont {Kitaura}}, \bibinfo {author} {\bibfnamefont
  {M.}~\bibnamefont {Yang}}, \ and\ \bibinfo {author} {\bibfnamefont
  {L.}~\bibnamefont {Li}},\ }\bibfield  {title} {\enquote {\bibinfo {title}
  {Dirac versus weyl fermions in topological insulators: Adler-bell-jackiw
  anomaly in transport phenomena},}\ }\href {\doibase
  10.1103/PhysRevLett.111.246603} {\bibfield  {journal} {\bibinfo  {journal}
  {Phys. Rev. Lett.}\ }\textbf {\bibinfo {volume} {111}},\ \bibinfo {pages}
  {246603} (\bibinfo {year} {2013})}\BibitemShut {NoStop}%
\bibitem [{\citenamefont {Huang}\ \emph {et~al.}(2015)\citenamefont {Huang},
  \citenamefont {Zhao}, \citenamefont {Long}, \citenamefont {Wang},
  \citenamefont {Chen}, \citenamefont {Yang}, \citenamefont {Liang},
  \citenamefont {Xue}, \citenamefont {Weng}, \citenamefont {Fang},
  \citenamefont {Dai},\ and\ \citenamefont {Chen}}]{huang2015}%
  \BibitemOpen
  \bibfield  {author} {\bibinfo {author} {\bibfnamefont {Xiaochun}\
  \bibnamefont {Huang}}, \bibinfo {author} {\bibfnamefont {Lingxiao}\
  \bibnamefont {Zhao}}, \bibinfo {author} {\bibfnamefont {Yujia}\ \bibnamefont
  {Long}}, \bibinfo {author} {\bibfnamefont {Peipei}\ \bibnamefont {Wang}},
  \bibinfo {author} {\bibfnamefont {Dong}\ \bibnamefont {Chen}}, \bibinfo
  {author} {\bibfnamefont {Zhanhai}\ \bibnamefont {Yang}}, \bibinfo {author}
  {\bibfnamefont {Hui}\ \bibnamefont {Liang}}, \bibinfo {author} {\bibfnamefont
  {Mianqi}\ \bibnamefont {Xue}}, \bibinfo {author} {\bibfnamefont {Hongming}\
  \bibnamefont {Weng}}, \bibinfo {author} {\bibfnamefont {Zhong}\ \bibnamefont
  {Fang}}, \bibinfo {author} {\bibfnamefont {Xi}~\bibnamefont {Dai}}, \ and\
  \bibinfo {author} {\bibfnamefont {Genfu}\ \bibnamefont {Chen}},\ }\bibfield
  {title} {\enquote {\bibinfo {title} {Observation of the
  chiral-anomaly-induced negative magnetoresistance in 3d weyl semimetal
  taas},}\ }\href {\doibase 10.1103/PhysRevX.5.031023} {\bibfield  {journal}
  {\bibinfo  {journal} {Phys. Rev. X}\ }\textbf {\bibinfo {volume} {5}},\
  \bibinfo {pages} {031023} (\bibinfo {year} {2015})}\BibitemShut {NoStop}%
\bibitem [{\citenamefont {Xiong}\ \emph {et~al.}(2015)\citenamefont {Xiong},
  \citenamefont {Kushwaha}, \citenamefont {Liang}, \citenamefont {Krizan},
  \citenamefont {Hirschberger}, \citenamefont {Wang}, \citenamefont {Cava},\
  and\ \citenamefont {Ong}}]{ong2015}%
  \BibitemOpen
  \bibfield  {author} {\bibinfo {author} {\bibfnamefont {Jun}\ \bibnamefont
  {Xiong}}, \bibinfo {author} {\bibfnamefont {Satya~K.}\ \bibnamefont
  {Kushwaha}}, \bibinfo {author} {\bibfnamefont {Tian}\ \bibnamefont {Liang}},
  \bibinfo {author} {\bibfnamefont {Jason~W.}\ \bibnamefont {Krizan}}, \bibinfo
  {author} {\bibfnamefont {Max}\ \bibnamefont {Hirschberger}}, \bibinfo
  {author} {\bibfnamefont {Wudi}\ \bibnamefont {Wang}}, \bibinfo {author}
  {\bibfnamefont {R.~J.}\ \bibnamefont {Cava}}, \ and\ \bibinfo {author}
  {\bibfnamefont {N.~P.}\ \bibnamefont {Ong}},\ }\bibfield  {title} {\enquote
  {\bibinfo {title} {Evidence for the chiral anomaly in the dirac semimetal
  na3bi},}\ }\href {\doibase 10.1126/science.aac6089} {\bibfield  {journal}
  {\bibinfo  {journal} {Science}\ }\textbf {\bibinfo {volume} {350}},\ \bibinfo
  {pages} {413--416} (\bibinfo {year} {2015})}\BibitemShut {NoStop}%
\bibitem [{\citenamefont {Li}\ \emph {et~al.}(2016)\citenamefont {Li},
  \citenamefont {Kharzeev}, \citenamefont {Zhang}, \citenamefont {Huang},
  \citenamefont {Pletikosic}, \citenamefont {Fedorov}, \citenamefont {Zhong},
  \citenamefont {Schneeloch}, \citenamefont {Gu},\ and\ \citenamefont
  {Valla}}]{valla2016}%
  \BibitemOpen
  \bibfield  {author} {\bibinfo {author} {\bibfnamefont {Qiang}\ \bibnamefont
  {Li}}, \bibinfo {author} {\bibfnamefont {Dmitri~E.}\ \bibnamefont
  {Kharzeev}}, \bibinfo {author} {\bibfnamefont {Cheng}\ \bibnamefont {Zhang}},
  \bibinfo {author} {\bibfnamefont {Yuan}\ \bibnamefont {Huang}}, \bibinfo
  {author} {\bibfnamefont {I.}~\bibnamefont {Pletikosic}}, \bibinfo {author}
  {\bibfnamefont {A.~V.}\ \bibnamefont {Fedorov}}, \bibinfo {author}
  {\bibfnamefont {R.~D.}\ \bibnamefont {Zhong}}, \bibinfo {author}
  {\bibfnamefont {J.~A.}\ \bibnamefont {Schneeloch}}, \bibinfo {author}
  {\bibfnamefont {G.~D.}\ \bibnamefont {Gu}}, \ and\ \bibinfo {author}
  {\bibfnamefont {T.}~\bibnamefont {Valla}},\ }\bibfield  {title} {\enquote
  {\bibinfo {title} {Chiral magnetic effect in zrte5},}\ }\href@noop {}
  {\bibfield  {journal} {\bibinfo  {journal} {Nat Phys}\ }\textbf {\bibinfo
  {volume} {12}},\ \bibinfo {pages} {550} (\bibinfo {year} {2016})}\BibitemShut
  {NoStop}%
\bibitem [{\citenamefont {Zhang}\ \emph {et~al.}(2016)\citenamefont {Zhang},
  \citenamefont {Xu}, \citenamefont {Belopolski}, \citenamefont {Yuan},
  \citenamefont {Lin}, \citenamefont {Tong}, \citenamefont {Bian},
  \citenamefont {Alidoust}, \citenamefont {Lee}, \citenamefont {Huang},
  \citenamefont {Chang}, \citenamefont {Chang}, \citenamefont {Hsu},
  \citenamefont {Jeng}, \citenamefont {Neupane}, \citenamefont {Sanchez},
  \citenamefont {Zheng}, \citenamefont {Wang}, \citenamefont {Lin},
  \citenamefont {Zhang}, \citenamefont {Lu}, \citenamefont {Shen},
  \citenamefont {Neupert}, \citenamefont {Zahid~Hasan},\ and\ \citenamefont
  {Jia}}]{jia2016}%
  \BibitemOpen
  \bibfield  {author} {\bibinfo {author} {\bibfnamefont {Cheng-Long}\
  \bibnamefont {Zhang}}, \bibinfo {author} {\bibfnamefont {Su-Yang}\
  \bibnamefont {Xu}}, \bibinfo {author} {\bibfnamefont {Ilya}\ \bibnamefont
  {Belopolski}}, \bibinfo {author} {\bibfnamefont {Zhujun}\ \bibnamefont
  {Yuan}}, \bibinfo {author} {\bibfnamefont {Ziquan}\ \bibnamefont {Lin}},
  \bibinfo {author} {\bibfnamefont {Bingbing}\ \bibnamefont {Tong}}, \bibinfo
  {author} {\bibfnamefont {Guang}\ \bibnamefont {Bian}}, \bibinfo {author}
  {\bibfnamefont {Nasser}\ \bibnamefont {Alidoust}}, \bibinfo {author}
  {\bibfnamefont {Chi-Cheng}\ \bibnamefont {Lee}}, \bibinfo {author}
  {\bibfnamefont {Shin-Ming}\ \bibnamefont {Huang}}, \bibinfo {author}
  {\bibfnamefont {Tay-Rong}\ \bibnamefont {Chang}}, \bibinfo {author}
  {\bibfnamefont {Guoqing}\ \bibnamefont {Chang}}, \bibinfo {author}
  {\bibfnamefont {Chuang-Han}\ \bibnamefont {Hsu}}, \bibinfo {author}
  {\bibfnamefont {Horng-Tay}\ \bibnamefont {Jeng}}, \bibinfo {author}
  {\bibfnamefont {Madhab}\ \bibnamefont {Neupane}}, \bibinfo {author}
  {\bibfnamefont {Daniel~S.}\ \bibnamefont {Sanchez}}, \bibinfo {author}
  {\bibfnamefont {Hao}\ \bibnamefont {Zheng}}, \bibinfo {author} {\bibfnamefont
  {Junfeng}\ \bibnamefont {Wang}}, \bibinfo {author} {\bibfnamefont {Hsin}\
  \bibnamefont {Lin}}, \bibinfo {author} {\bibfnamefont {Chi}\ \bibnamefont
  {Zhang}}, \bibinfo {author} {\bibfnamefont {Hai-Zhou}\ \bibnamefont {Lu}},
  \bibinfo {author} {\bibfnamefont {Shun-Qing}\ \bibnamefont {Shen}}, \bibinfo
  {author} {\bibfnamefont {Titus}\ \bibnamefont {Neupert}}, \bibinfo {author}
  {\bibfnamefont {M.}~\bibnamefont {Zahid~Hasan}}, \ and\ \bibinfo {author}
  {\bibfnamefont {Shuang}\ \bibnamefont {Jia}},\ }\bibfield  {title} {\enquote
  {\bibinfo {title} {Signatures of the adler-bell-jackiw chiral anomaly in a
  weyl fermion semimetal},}\ }\href@noop {} {\bibfield  {journal} {\bibinfo
  {journal} {Nat Commun}\ }\textbf {\bibinfo {volume} {7}},\ \bibinfo {pages}
  {10735} (\bibinfo {year} {2016})}\BibitemShut {NoStop}%
\bibitem [{\citenamefont {Volovik}(2003)}]{volovik2003}%
  \BibitemOpen
  \bibfield  {author} {\bibinfo {author} {\bibfnamefont {Grigorij~E}\
  \bibnamefont {Volovik}},\ }\href@noop {} {\emph {\bibinfo {title} {The
  universe in a helium droplet}}}\ (\bibinfo  {publisher} {Oxford},\ \bibinfo
  {year} {2003})\BibitemShut {NoStop}%
\bibitem [{\citenamefont {{Liu}}\ \emph {et~al.}(2016)\citenamefont {{Liu}},
  \citenamefont {{Fang}},\ and\ \citenamefont {{Fu}}}]{liu2016}%
  \BibitemOpen
  \bibfield  {author} {\bibinfo {author} {\bibfnamefont {J.}~\bibnamefont
  {{Liu}}}, \bibinfo {author} {\bibfnamefont {C.}~\bibnamefont {{Fang}}}, \
  and\ \bibinfo {author} {\bibfnamefont {L.}~\bibnamefont {{Fu}}},\ }\bibfield
  {title} {\enquote {\bibinfo {title} {{Tunable Weyl fermions and Fermi arcs in
  magnetized topological crystalline insulators}},}\ }\href@noop {} {\bibfield
  {journal} {\bibinfo  {journal} {ArXiv e-prints}\ } (\bibinfo {year}
  {2016})},\ \Eprint {http://arxiv.org/abs/1604.03947} {arXiv:1604.03947
  [cond-mat.mes-hall]} \BibitemShut {NoStop}%
\bibitem [{\citenamefont {{Tang}}\ \emph {et~al.}(2016)\citenamefont {{Tang}},
  \citenamefont {{Zhou}}, \citenamefont {{Xu}},\ and\ \citenamefont
  {{Zhang}}}]{tang2016}%
  \BibitemOpen
  \bibfield  {author} {\bibinfo {author} {\bibfnamefont {P.}~\bibnamefont
  {{Tang}}}, \bibinfo {author} {\bibfnamefont {Q.}~\bibnamefont {{Zhou}}},
  \bibinfo {author} {\bibfnamefont {G.}~\bibnamefont {{Xu}}}, \ and\ \bibinfo
  {author} {\bibfnamefont {S.-C.}\ \bibnamefont {{Zhang}}},\ }\bibfield
  {title} {\enquote {\bibinfo {title} {{Dirac Fermions in Antiferromagnetic
  Semimetal}},}\ }\href@noop {} {\bibfield  {journal} {\bibinfo  {journal}
  {ArXiv e-prints}\ } (\bibinfo {year} {2016})},\ \Eprint
  {http://arxiv.org/abs/1603.08060} {arXiv:1603.08060 [cond-mat.mtrl-sci]}
  \BibitemShut {NoStop}%
\bibitem [{\citenamefont {Vazifeh}\ and\ \citenamefont
  {Franz}(2013)}]{vazifeh2013}%
  \BibitemOpen
  \bibfield  {author} {\bibinfo {author} {\bibfnamefont {M.~M.}\ \bibnamefont
  {Vazifeh}}\ and\ \bibinfo {author} {\bibfnamefont {M.}~\bibnamefont
  {Franz}},\ }\bibfield  {title} {\enquote {\bibinfo {title} {Electromagnetic
  response of weyl semimetals},}\ }\href {\doibase
  10.1103/PhysRevLett.111.027201} {\bibfield  {journal} {\bibinfo  {journal}
  {Phys. Rev. Lett.}\ }\textbf {\bibinfo {volume} {111}},\ \bibinfo {pages}
  {027201} (\bibinfo {year} {2013})}\BibitemShut {NoStop}%
\bibitem [{\citenamefont {Abrikosov}(1972)}]{abrikosov}%
  \BibitemOpen
  \bibfield  {author} {\bibinfo {author} {\bibfnamefont {A.~A.}\ \bibnamefont
  {Abrikosov}},\ }\href@noop {} {\emph {\bibinfo {title} {Introduction to the
  theory of normal metals}}}\ (\bibinfo  {publisher} {Academic Press, New
  York},\ \bibinfo {year} {1972})\BibitemShut {NoStop}%
\bibitem [{\citenamefont {Wang}\ \emph {et~al.}(2007)\citenamefont {Wang},
  \citenamefont {Xu}, \citenamefont {Shimono}, \citenamefont {Tanaka},\ and\
  \citenamefont {Yamazaki}}]{2007MAW200717}%
  \BibitemOpen
  \bibfield  {author} {\bibinfo {author} {\bibfnamefont {Haitao}\ \bibnamefont
  {Wang}}, \bibinfo {author} {\bibfnamefont {Yibin}\ \bibnamefont {Xu}},
  \bibinfo {author} {\bibfnamefont {Masato}\ \bibnamefont {Shimono}}, \bibinfo
  {author} {\bibfnamefont {Yoshihisa}\ \bibnamefont {Tanaka}}, \ and\ \bibinfo
  {author} {\bibfnamefont {Masayoshi}\ \bibnamefont {Yamazaki}},\ }\bibfield
  {title} {\enquote {\bibinfo {title} {Computation of interfacial thermal
  resistance by phonon diffuse mismatch model},}\ }\href {\doibase
  10.2320/matertrans.MAW200717} {\bibfield  {journal} {\bibinfo  {journal}
  {Mater, trans.}\ }\textbf {\bibinfo {volume} {48}},\ \bibinfo {pages}
  {2349--2352} (\bibinfo {year} {2007})}\BibitemShut {NoStop}%
\bibitem [{\citenamefont {Bonn}\ and\ \citenamefont {Hardy}(2016)}]{note1}%
  \BibitemOpen
  \bibfield  {author} {\bibinfo {author} {\bibfnamefont {D.A.}\ \bibnamefont
  {Bonn}}\ and\ \bibinfo {author} {\bibfnamefont {W.N.}\ \bibnamefont
  {Hardy}},\ }\href@noop {} {\  (\bibinfo {year} {2016})},\ \bibinfo {note}
  {private communication.}\BibitemShut {Stop}%
\bibitem [{\citenamefont {Sumiyoshi}\ and\ \citenamefont
  {Fujimoto}(2016)}]{fujimoto2016}%
  \BibitemOpen
  \bibfield  {author} {\bibinfo {author} {\bibfnamefont {Hiroaki}\ \bibnamefont
  {Sumiyoshi}}\ and\ \bibinfo {author} {\bibfnamefont {Satoshi}\ \bibnamefont
  {Fujimoto}},\ }\bibfield  {title} {\enquote {\bibinfo {title} {Torsional
  chiral magnetic effect in a weyl semimetal with a topological defect},}\
  }\href {\doibase 10.1103/PhysRevLett.116.166601} {\bibfield  {journal}
  {\bibinfo  {journal} {Phys. Rev. Lett.}\ }\textbf {\bibinfo {volume} {116}},\
  \bibinfo {pages} {166601} (\bibinfo {year} {2016})}\BibitemShut {NoStop}%
\bibitem [{\citenamefont {Schuster}\ \emph {et~al.}(2016)\citenamefont
  {Schuster}, \citenamefont {Iadecola}, \citenamefont {Chamon}, \citenamefont
  {Jackiw},\ and\ \citenamefont {Pi}}]{chamon2016}%
  \BibitemOpen
  \bibfield  {author} {\bibinfo {author} {\bibfnamefont {Thomas}\ \bibnamefont
  {Schuster}}, \bibinfo {author} {\bibfnamefont {Thomas}\ \bibnamefont
  {Iadecola}}, \bibinfo {author} {\bibfnamefont {Claudio}\ \bibnamefont
  {Chamon}}, \bibinfo {author} {\bibfnamefont {Roman}\ \bibnamefont {Jackiw}},
  \ and\ \bibinfo {author} {\bibfnamefont {So-Young}\ \bibnamefont {Pi}},\
  }\bibfield  {title} {\enquote {\bibinfo {title} {Dissipationless conductance
  in a topological coaxial cable},}\ }\href {\doibase
  10.1103/PhysRevB.94.115110} {\bibfield  {journal} {\bibinfo  {journal} {Phys.
  Rev. B}\ }\textbf {\bibinfo {volume} {94}},\ \bibinfo {pages} {115110}
  (\bibinfo {year} {2016})}\BibitemShut {NoStop}%
\bibitem [{\citenamefont {Groth}\ \emph {et~al.}(2014)\citenamefont {Groth},
  \citenamefont {Wimmer}, \citenamefont {Akhmerov},\ and\ \citenamefont
  {Waintal}}]{groth14}%
  \BibitemOpen
  \bibfield  {author} {\bibinfo {author} {\bibfnamefont {Christoph~W}\
  \bibnamefont {Groth}}, \bibinfo {author} {\bibfnamefont {Michael}\
  \bibnamefont {Wimmer}}, \bibinfo {author} {\bibfnamefont {Anton~R}\
  \bibnamefont {Akhmerov}}, \ and\ \bibinfo {author} {\bibfnamefont {Xavier}\
  \bibnamefont {Waintal}},\ }\bibfield  {title} {\enquote {\bibinfo {title}
  {Kwant: a software package for quantum transport},}\ }\href@noop {}
  {\bibfield  {journal} {\bibinfo  {journal} {New Journal of Physics}\ }\textbf
  {\bibinfo {volume} {16}},\ \bibinfo {pages} {063065} (\bibinfo {year}
  {2014})}\BibitemShut {NoStop}%
\bibitem [{\citenamefont {Cano}\ \emph {et~al.}(unpublished)\citenamefont
  {Cano}, \citenamefont {Bradlyn}, \citenamefont {Wang}, \citenamefont
  {Hirschberger}, \citenamefont {Ong},\ and\ \citenamefont
  {Bernevig}}]{cano16}%
  \BibitemOpen
  \bibfield  {author} {\bibinfo {author} {\bibfnamefont {Jennifer}\
  \bibnamefont {Cano}}, \bibinfo {author} {\bibfnamefont {Barry}\ \bibnamefont
  {Bradlyn}}, \bibinfo {author} {\bibfnamefont {Zhijun}\ \bibnamefont {Wang}},
  \bibinfo {author} {\bibfnamefont {Max}\ \bibnamefont {Hirschberger}},
  \bibinfo {author} {\bibfnamefont {NP}~\bibnamefont {Ong}}, \ and\ \bibinfo
  {author} {\bibfnamefont {BA}~\bibnamefont {Bernevig}},\ }\bibfield  {title}
  {\enquote {\bibinfo {title} {The chiral anomaly factory: Creating weyls with
  a magnetic field},}\ }\href@noop {} {\bibfield  {journal} {\bibinfo
  {journal} {arXiv:1604.08601}\ } (\bibinfo {year} {unpublished})}\BibitemShut
  {NoStop}%
\end{thebibliography}%

%%%%%%%%%%%%%%%%%%%%%%%%%%%%%%%%%%%%%%%%%%%%%%%%%%

%\newpage

\appendix

\section{Tight binding model, dispersion relations and parameters for Cd$_3$As$_2$ and Na$_3$Bi}
\begin{figure*}[t]
\includegraphics[width = 17.5cm]{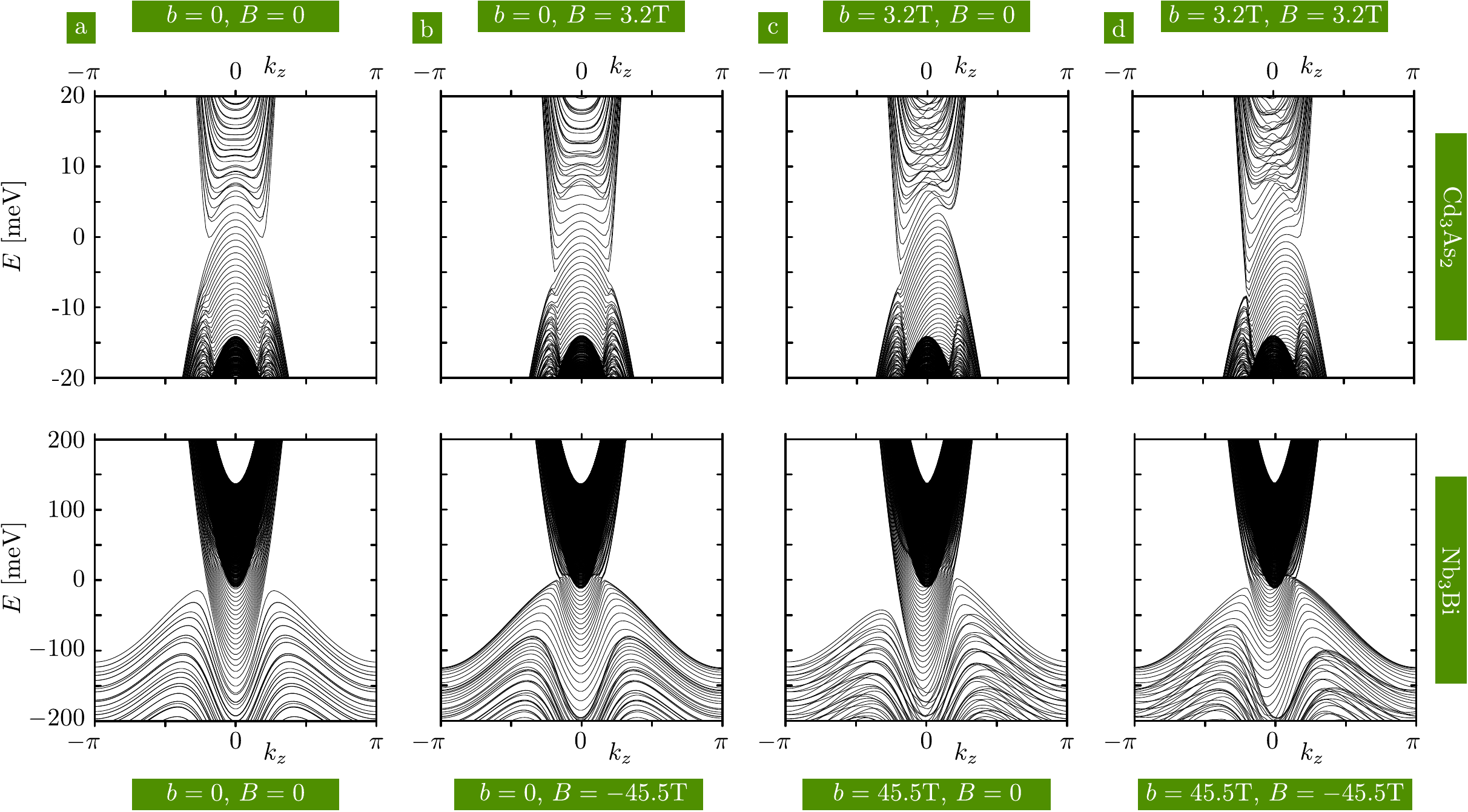}
\caption{Dispersion relations for the spin-up sector of the lattice Hamiltonian (\ref{h03}) describing  Cd$_3$As$_2$ (top row) and Na$_3$Bi (bottom row). The parameters used in the simulations include the particle-hole symmetry breaking terms and are summarized in Table I. We used a lattice with $40\times 40$ sites and the magnetic fields shown in the green boxes for each of the material. Notice the different magnitude of effective magnetic fields for different compounds -- this is due to the different lattice constants, and different sign of the physical magnetic field between the  two rows. Different sign of magnetic fields shows that the physical magnetic field compensates the torsional one in opposite Weyl points for opposite directions of magnetic field in accordance with the interpretation in the main text.
}\label{fig1_sup}
\end{figure*}

From the low energy $\bk\cdot\bp$ Hamiltonian (\ref{h2}) we construct the requisite lattice model by replacing $Ak_\pm\to (A/a)(\sin{ak_x}\pm i\sin{ak_y})$, $C_1 k_z^2\to (2C_1/a^2)(1-\cos{ak_z})$, etc. For example $\epsilon_0(\bk)$ defined below Eq.\ (\ref{h2}) becomes
\bee\label{ap1}
\epsilon_\bk=c_0+c_1\cos{ak_z} +c_2(\cos{ak_x}+\cos{ak_y})
\ee
with $c_0=C_0+2(C_1+2C_2)/a^2$, $c_1=-2C_1/a^2$ and $c_2=-2C_2/a^2$. 
The constants $c_j$ are chosen such that $\epsilon_\bk$ matches $\epsilon_0(\bk)$ for small $ak$ independent of the chosen value of the lattice constant $a$. 
Treating all other terms in the Hamiltonian (\ref{h2}) in the similar fashion leads to the lattice Hamiltonian given by Eqs. (\ref{h03}) and (\ref{h3}). 

In addition to the results presented in the main text we performed detailed bandstructure simulations for Dirac semimetals Cd$_3$As$_2$ and Na$_3$Bi. Parameters for the model Hamiltonian \eqref{h2} are taken from \cite{zhizhun2013,cano16}  and \citep{zhizhun2012} correspondingly and are summarized in the Table \ref{table1}. In the main text we only presented results for the parameters of Cd$_3$As$_2$ with the asymmetry parameters $C_i$ set to zero. In the Fig.\ \ref{fig1_sup} we present the dispersion relation computation for the models of $\frac{1}{2}$-Cd$_3$As$_2$ and $\frac{1}{2}$-Na$_3$Bi with all the asymmetry terms taken into account. The effects discussed in the main text are present even in this more general case although to see them clearly now requires more effort due to the more complicated structure of the energy bands.  For instance the equivalence of the torsional strain and magnetic field, pointed out in the main text, here can be only identified by a trained eye. One needs to notice that the right Weyl point is at $E=0$ in the rightmost graph of the first column of Fig.\ \ref{fig1_sup}. Results between the two parameter sets are similar, but notice the larger gaps in Na$_3$Bi, which may make the experimental realization easier.
\begin{table}\label{table1}
% \begin{tabular}{l | c | c }
\begin{tabular*}{0.48\textwidth}{@{\extracolsep{\fill}}lcc}
\hline \hline 
 & \   Cd$_3$As$_2$ \  & \  Na$_3$Bi \  \\
\hline
$C_0$ [eV] & -0.0145 & -0.0638 \\
$C_1$ [eV\AA$^2$] & 10.59 & 8.75\\
$C_2$ [eV\AA$^2$] & 11.5 & -8.4 \\
$M_0$ [eV] & 0.0205 & 0.869 \\
$M_1$ [eV\AA$^2$] & -18.77 & -10.64\\
$M_2$ [eV\AA$^2$] & -13.5 & -10.36 \\
$A$ [eV\AA] & 0.889 &  2.46 \\
$a$ [\AA]  & 20 & 7.5 \\
\hline \hline
\end{tabular*}
\caption{Material parameters taken from \citep{zhizhun2012} and \cite{cano16} used for our simulations. The last row represents the effective lattice constant used for simulations in Fig.\ \ref{fig1_sup}.}
\end{table}

\begin{figure}[b]
\includegraphics[width = 1.0\linewidth]{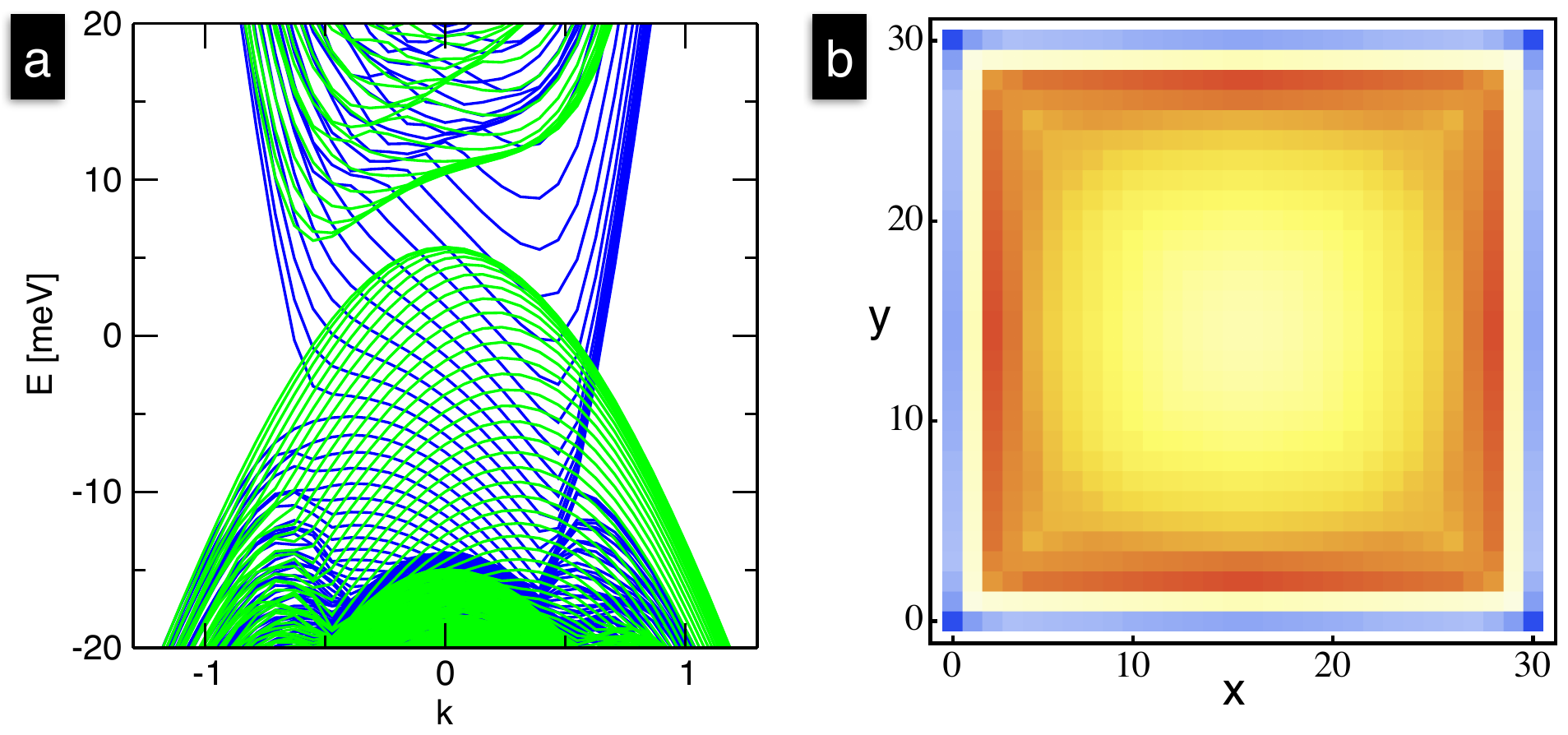}
\caption{Persistent currents in a Cd$_3$As$_2$ nanowire under torsion and magnetic field. a) Band structure detail for spin up (blue) and spin down (green) sectors in a $30\times 30$ lattice with $B=b=3.2$T and other parameters as in Fig.\ \ref{fig1_sup}.  b) Calculated current density $j_z$ for $\mu=0$ including contributions from both spin sectors.}
\label{fig3_supp}
\end{figure}

To further confirm the validity of our ideas relating the torsional strain to the pseudomagnetic field we computed the equilibrium currents flowing along the wire using the full Hamiltonian including the p-h symmetry breaking terms. The results are as follows: (i) For both  Cd$_3$As$_2$ and Na$_3$Bi we find equilibrium currents with the pattern similar to the one displayed in Fig.\ \ref{fig3} in each spin sector when nonzero torsion is present. The total current density (summing up contributions from both spin up and down  sectors) vanishes, as it must be in a $\cT$-invariant system.   (ii) When only magnetic field $B$ is present and no torsion, the current densities are zero in both sectors separately, in accord with the expectation. (In this case the band structure in each spin sector shows the same number of left and right moving modes in the bulk of the system). (iii) When both torsion and magnetic field are present then we find non-zero persistent current density in both spin sectors. In this case $\cT$ is absent and the currents from the two sectors  generically do not cancel. This is illustrated in Fig.\ \ref{fig3_supp}. We observe an asymmetric band structure that supports different number of bulk left and right moving  modes at various energies, leading to a net imbalance in the current flow between the bulk and the surface. The total current carried by the wire however still vanishes. 

%%%%
\section{Nonequilibrium distribution in a stretched system}
In this Appendix we derive a quantitative estimate for the chemical potential denoted by $\mu'_{\rm eq}$ in Fig.\ \ref{fig8} as well as the corresponding bulk electron density. As discussed in Sec.\ III.B upon introducing strain the chemical potential in the bulk of the system rises from $\mu_0$ to $\mu'$ while that at the surfaces remains unchanged. In the process, the charge density also remains unchanged (except possibly for perturbations near the surface that average to zero and do not affect the bulk). This creates a nonequilibrium distribution of electrons illustrated in Fig.\ \ref{fig8}b which then relaxes to a new equilibrium characterized by a global chemical potential $\mu'_{\rm eq}$. The latter can be calculated by demanding that the total electron number $N$ is conserved. 

In the unstrained system we have $N=N_s+N_b$ where the subscripts refer to the surface and the bulk, respectively. In the nonequilibrium strained system $N_s$ and $N_b$ remain the same as per our discussion above. In the new equilibrium they change to $N_{s/b}'= N_{s/b}+\delta N_{s/b}$, where $\delta N_{s/b}\simeq\kappa_{s/b}\delta\mu_{s/b}$. Here $\kappa_{s/b}=dN_{s/b}/d\mu$ is the compressibility and $\delta\mu_{s/b}$ denotes the change in the chemical potential that is responsible for the change in $N_{s/b}$. Number conservation dictates that $\delta N_s=-\delta N_b$ which implies
\bee\label{b1}
\kappa_s(\mu'_{\rm eq}-\mu_0)=-\kappa_b(\mu'_{\rm eq}-\mu').
\ee
We can solve for $\mu'_{\rm eq}$ to obtain
\bee\label{b2}
\mu'_{\rm eq}=\mu_0+{\delta\mu\over 1+\kappa_s/\kappa_b}, 
\ee
where  $\delta\mu=\mu'-\mu_0$. The corresponding change in the bulk number is $\delta N_b=\kappa_b(\mu'_{\rm eq}-\mu')$ which, together with Eq.\ (\ref{b2})  gives the change in the bulk density 
\bee\label{b3}
\delta\rho^{\rm bulk}=-{1\over Sd}{\kappa_b\over 1+\kappa_b/\kappa_s}\delta\mu,
\ee
where $S$ is the area of the slab.

To complete the calculation we require the surface and bulk electron compressibilities. For the surface we assume that we have  a single linearly dispersing band $\epsilon^s_\bk=\hbar v\ k_x$ on each surface that extends between the surface projections of the two Weyl points, $|k_z|<Q$. We furthermore assume that the surface state is essentially unaffected by the magnetic field, in accord with the results of out lattice simulations. This gives
\bee\label{b4}
\kappa_s=SD_s(\mu)={SQ\over \pi^2\hbar v},
\ee
where $D_s(\epsilon)=Q/\pi^2\hbar v$ is the surface density of states (counting both surfaces). 

For the bulk we similarly have $\kappa_b=SdD_b(\mu)$. We now must distinguish between the quantum and the semiclassical limits, as defined in Sec.\ IV.B.  In the quantum limit we have a pair of linearly dispersing $n=0$ Dirac Landau levels with degeneracy $(B/\Phi_0)$ whereas in the semiclassical limit many Landau levels are populated so it is permissible to approximate the density of states by that of a zero-field system. We thus obtain
\bee\label{b5}
\kappa_b=\left\{
\begin{array}{ll}
{Sd\over\pi\hbar v}\left({B\over \Phi_0}\right) \  & {\rm quantum \ limit}, \\
{Sd\mu^2\over \pi^2\hbar^3 v^3}\ & {\rm semiclassical \ limit}.
\end{array}
\right.
\ee
Substituting these results into Eqs.\ (\ref{b2}) and (\ref{b3}) we obtain results for  
$\mu'_{\rm eq}$ and $\delta\rho^{\rm bulk}$ quoted in the main text (Eqs.\ \ref{p2} and \ref{p3}) for the quantum limit. In the semiclassical limit we similarly obtain
\bee\label{pp2}
\mu'_{\rm eq}=\mu_0+{\delta\mu\over 1+\lambda_Q/d}, 
\ee
and
\bee\label{pp3}
\delta\rho^{\rm bulk}=-{\alpha\over \pi a}\left({B\over \Phi_0}\right){\cot{aQ}\over 1+d/\lambda_Q},
\ee
where $\lambda_Q=Q(\hbar v/\mu)^2/2$ is the characteristic length.

%%%%
\section{Hydrodynamic flow in a twisted Weyl nanowire}

Consider a cylindrical nanowire of radius $R$ made of a Weyl semimetal. Both torsion and electric field $\bE$ are applied along the axis of the wire (taken here along the $\hat{z}$ direction), giving a non-zero right hand side $\propto\bb\cdot\bE$ in the second anomaly equation (\ref{an2}). Denoting the right hand side by $g(r)$ we may write 
\bee\label{a1}
\partial_t\rho+\nabla\cdot\bj=g(r),
\ee
where
\bee\label{a2}
g(r)=g_0\left[\theta(R-r)-{1\over 2}R\delta(r-R)\right], 
\ee
with $g_0=(e^2/ 2\pi^2\hbar^2 c)bE$. The first term in $g(r)$ describes uniform production of electrons in the bulk of the wire at a rate given by the chiral anomaly. The second term reflects the fact that those electrons are removed from the surface, in accord with our discussion in Section III. The total production in the wire is zero, $\int_0^{R+}r dr g(r)=0$, and the charge is conserved.

We now assume that the dominant relaxation mechanism for the nonequilibrium electrons produced in the bulk of the wire is diffusion towards the boundary. Electrons move ballistically along the $\hat{z}$ direction and undergo occasional  collisions that scatter them into neighboring Landau level states. Near the boundary bulk electrons can finally backscatter into the surface modes which are moving in the opposite direction. Under this assumption the diffusion current is 
\bee\label{a3}
\bj=-D\nabla\rho
\ee
where $D=\ell_b^2/\tau_0$ is the diffusion constant ($\ell_b=\sqrt{\hbar c/eb}$ is the magnetic length and $\tau_0^{-1}$ the microscopic scattering rate). The form of the diffusion constant reflects the fact that electron wavefunctions have Landau level character with the spatial extent $\ell_b$ in the direction perpendicular to $\hat{z}$ and 
scattering occurs predominantly between neighboring Landau level orbitals.

Substituting Eq.\ (\ref{a3}) into (\ref{a1}) and specializing to long time steady state with $\partial_t\rho=0$ we obtain 
\bee\label{a4}
-D\nabla^2\rho=g(r).
\ee
Writing the Laplacian in the polar coordinates and assuming radially symmetric solution we find 
\bee\label{a5}
\rho(r)={g_0\over D}\left({R^2-r^2\over 4}\right)\theta(R-r). 
\ee
The corresponding radial diffusion current density is $j_\br(r)=-D\partial_r\rho={1\over 2}g_0r$. The total non-equilibrium charge in the bulk modes is
\bee\label{a6}
\delta Q=-e\int_0^Rdr 2\pi r\rho(r)=-e{\pi\over 8} {g_0\over D}R^4.
\ee
Since all these modes move in the same direction with velocity $v$ this gives the total current along the $\hat{z}$ direction in the wire
\bee\label{a7}
J_{\rm CTE}=2v\delta Q=-ev{\pi\over 4}\left({e^2\over 2\pi^2\hbar^2 c}\right){\tau\over\ell_b^2}R^4bE.
\ee
The factor 2 in the first equality reflects the fact that non-equilibrium charge $-\delta Q$ must exist in the surface left moving modes to maintain overall charge neutrality. We assume for simplicity that these modes move at the same speed $v$.  
\begin{figure}[b]
\includegraphics[width = 0.9\linewidth]{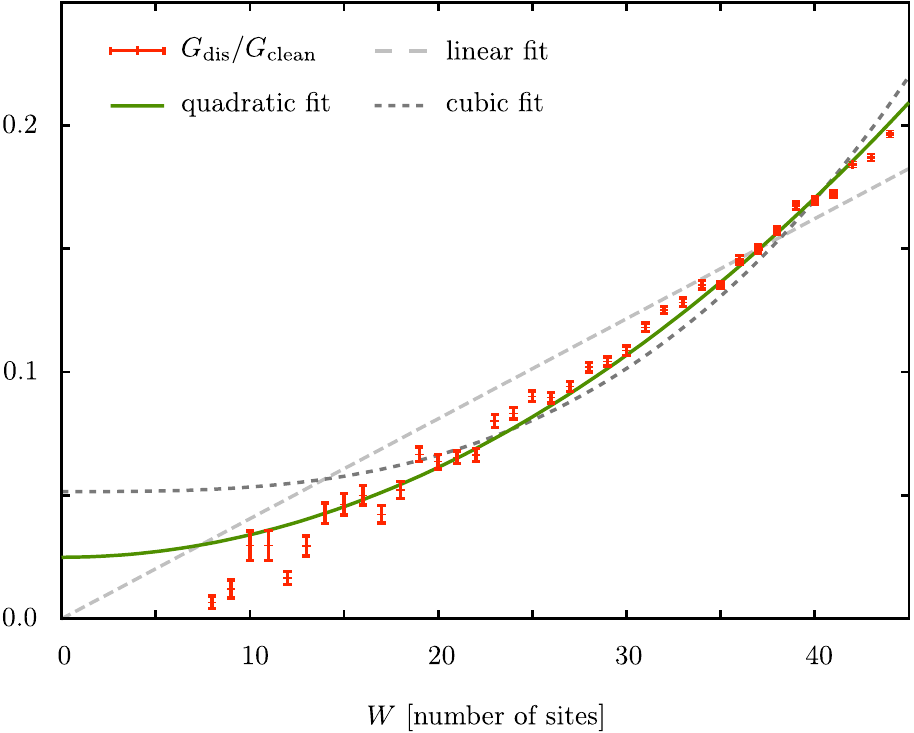}
\caption{Ratio of conductance of a disordered $W\times W \times 20$ system to the conductance of the clean system averaged over $100$ disorder realizations. Green line is the best fit to the data -- parabolic, grey curves show the failure of the linear (with non-negative $G(0)$) and cubic fits.
}\label{fig2_supp}
\end{figure}

As mentioned, the transport current along the wire exhibits all the characteristics of the hydrodynamic flow: it is largest at the center and vanishes at the boundary. This is because momentum can only be relaxed by electrons that have reached the boundary and can scatter into surface modes. The amount of current through the wire scales with $R^4$, just like fluid flowing through a pipe.

From Eq.\ (\ref{a7}) one can read off the chiral torsional conductivity $\sigma_{\rm CTE}$ which can be written suggestively in the following way
\bee\label{a8}
\sigma_{\rm CTE}={e^2v \over 4\pi h}\tau {\cal N},
\ee
where ${\cal N}=\pi R^2/\ell_b^2$ is the number of chiral bulk modes in the wire and $\tau=\tau_0R^2/\ell_b^2$ is the effective transport scattering time. The form of the latter reflects the fact that under diffusion the electron produced near the center of the wire must scatter on average  $(R/\ell_b)^2$ times before it reaches the boundary.

To illustrate this point we have performed simulation of conductance in a disordered symmetric $\frac{1}{2}$-Cd$_3$As$_2$ model. The Hamiltonian parameters are the same as used in Fig.\ \ref{fig2}. We have performed the conductance simulations for $\mu=5$meV and for the system of $W\times W \times 20$ sites. We have added on-site disorder $\delta \mu_i$ taken from normal distribution of width $10$meV to simulate the hydrodynamic flow described above. The ratio of conductance of disordered system to the conductance of the clean system is plotted in Fig.\ \ref{fig2_supp}. Best fit to the data is in accordance with \eqref{a8}, where $\tau \propto R^2$.

%%%%%%%%%%%%%%%%%%%%%%%%%%%%%%%%%%%%%%%%%%%%%%%%%%
\end{document}